\documentclass[preprint,authoryear,12pt,nodate]{modified}
\usepackage{tablefootnote}
\usepackage{graphicx}
\usepackage{amssymb}
\usepackage{amsthm}
\usepackage{mathtools}
\usepackage{color, colortbl}
\usepackage{datetime}
\definecolor{LightGray}{rgb}{0.82,0.82,0.82}
\newdate{date}{5}{7}{2014}

\journal{}

\begin{document}

\begin{frontmatter}

\title{Dynamic Coupling and Market Instability}

\renewcommand*{\thefootnote}{\fnsymbol{footnote}}

\author{Christopher D. Clack~\footnote{clack@cs.ucl.ac.uk.  Corresponding Author.}}
\author{Elias Court}%~\footnote{Elias.Court.09@ucl.ac.uk}}
\author{Dmitrijs Zaparanuks~\footnote{%zaparanuks@gmail.com.  
Author was supported by SNSF grant~\includegraphics[width = 10mm]{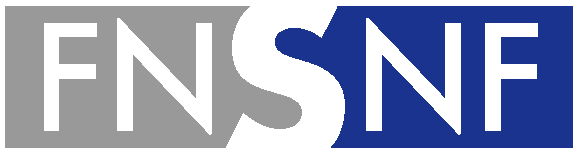}}\setcounter{footnote}{0}}
\address{University College London, Gower Street, London WC1E 6BT, UK
\break

5th July 2014}

\begin{abstract}
We examine dynamic coupling and feedback effects between High Frequency Traders (HFTs) and how they can destabilize markets.
We develop a general framework for modelling dynamic interaction based on recurrence relations, and use this to 
show how unexpected latency and feedback can trigger oscillatory instability between HFT market makers with inventory constraints.
Our analysis suggests that the modelled instability is an unintentional emergent behaviour of the market that does not depend on the complexity of  HFT strategies --- even apparently stable strategies are vulnerable.
Feedback instability can lead to substantial movements in market prices such as price spikes and crashes.
\end{abstract}

\begin{keyword}
%% keywords here, in the form: keyword \sep keyword
HFT \sep market maker \sep latency  \sep
feedback  \sep instability \sep coupling

%% MSC codes here, in the form: \MSC code \sep code
%% or \MSC[2008] code \sep code (2000 is the default)
\MSC C61 \sep  63 \sep  G19 \sep  G14
%C53 - Forecasting and Prediction Methods; Simulation Methods 
%C61 - Optimization Techniques; Programming Models; Dynamic Analysis 
%C63 - Computational Techniques; Simulation Modeling 
%G14 - Information and Market Efficiency; Event Studies
%G1 - General Financial Markets  
%G19 - Other 

\end{keyword}

\end{frontmatter}

\section{Introduction}
\label{introduction}
In a recent journal special issue on High Frequency Trading (HFT) Chordia et al highlighted a key unanswered problem: what is ``the nature of the mechanism by which the interaction of HFT algorithms improves market quality''? \citep{Chordia2013}.  In the same issue, Hasbrouck and Saar said they ``cannot rule out that in times of severe 
market conditions HFTs may contribute to market failure'' \citep{Hasbrouck2013}.  Here, we contribute to this debate by investigating the 
low-level mechanisms by which High Frequency Traders (HFTs) may interact to reduce market quality and lead to 
failure, especially during times of market disequilibrium.  
We show how coupling and feedback loops may occur between HFTs, we introduce a general framework for 
modelling dynamic interaction between financial algorithms, and we 
show how latency and feedback loops may trigger instability as an unintentional emergent behaviour of the market.\footnote{This follows a more pragmatic and sector-specific approach than our prior work on modelling emergent behaviour such as \cite{Chen2007,Chen2008,Chen2009,Chen2010}.}

Concern has previously been expressed about the potential for feedback loops to impact prices and 
destabilize markets \citep{Danielsson2012,Zigrand2012a}.
Feedback loops are not only widespread within the financial markets but may also exist for a long time, 
in some cases possibly remaining unnoticed.
The adverse effects
of a destabilising feedback loop may only become apparent when its strength 
becomes sufficiently large.

A prominent example of market instability either arising from or exacerbated by HFTs was the 
Flash Crash of May 6th 2010
\citep{CFTC-SEC2010a}.  
Of direct relevance to our work is the ``hot potato'' trading
behaviour of market makers at the heart of the Flash Crash, where multiple HFTs traded with each other in a 
rapid oscillation of large aggressive orders. This highly unusual oscillatory instability   
created both deceptive trading volume (which implied
liquidity where none was present) and a spike in messaging traffic that 
stressed
the already-overloaded technology 
infrastructure \citep{Nanex2010c}.  

We model dynamic coupling and feedback between HFTs at the level of the market microstructure, 
and expose the underlying mechanics of dynamic interaction.
We illustrate this with a case study of interaction between inventory-driven HFT market makers 
\citep{Menkveld2013}
in an order-book market, each executing a simple, stable trading strategy. 
Unlike those observed by \citep{Hasbrouck2013}, our HFTs do not intentionally interact or ``play'' with 
each other.
Nevertheless, we explain how dynamic coupling between our HFTs 
leads to a feedback loop where each HFT influences the 
behaviour of the other, and how this feedback has the potential to 
generate unintentional  
instability (including highly volatile oscillatory behaviour) as an emergent behaviour of the market.

Our model shows that one of the triggers for such behaviour is the introduction of unexpected additional latency 
(unexpected delay), as might be experienced for example when
a sudden burst of quotes overwhelms an execution venue and causes market data feeds to be delayed.  
We illustrate how we model and analyse a market where traders experience unexpected delay and how this 
leads to instability.   

The primary contribution of this article is to provide an alternative narrative for market instability, using our dynamic interaction model to show in great detail {\em how} latency and feedback loops between dynamically-coupled HFTs may trigger unintentional market instability.  Although our case study focuses on oscillation arising from the interaction of automated market-making strategies, 
we suspect that many of the previously observed feedback loops 
(e.g. in \citet{Danielsson2012}) and the impact of market maker inventories on time-varying liquidity \citep{Comerton-Forde2010} may also be modelled and analysed using the techniques
we describe in this article.  Our work may also have implications for models of pricing and market impact, since we demonstrate that traders do not necessarily have independence of action and such models may need to account for unexpected coupling with other traders.

The paper is organized as follows. First we set out the relationship with prior work, followed by an introduction to modelling coupling and feedback.  Section~\ref{sec:model} details our dynamic interaction model for a simple case study, and  
Section~\ref{sec:analysis} analyses this model and makes the link from coupling and feedback to market instability (including numerical simulation results, which illustrate some theoretically infinite modes of oscillation).  Section~\ref{sec:summary} concludes, and is followed by an appendix containing further definitions for our case-study.

\section{Relation to prior work}
\label{related}

Previous theoretical and empirical studies of instability
have predominantly focused on interaction as 
an indirect process via prices 
\citep{Arthur1996,Caldarelli1997},
or via globally-shared 
information \citep{
Brock1998,
Lux1999,
Hommes2009} without providing a detailed exploration of the underlying mechanism of interaction that causes such price fluctuations.  
Where direct interaction is included in the model, it is often abstracted --- for example, using an Ising model to provide an abstraction of nearest-neighbour communication between 
traders
\citep{Kaizoji2000,Iori2002b}, or assuming traders form bidirectional links in a random network
\citep{Cont2000}.
By contrast, we explore the mechanistic order-by-order interaction within the limit order book, where we explicitly model multiple bilateral direct interactions between traders.

We follow \citet{Iori2002b} and \citet{Cvitanic2010} by modelling dynamic interaction in discrete time, thereby exposing substantial microstructure detail such as the discrete nature of computer messaging, of order processing by heterogeneous traders, and the discrete nature of order arrival and order processing by the limit order book \citep{Day1990}.

We have found recurrence relations to be the most helpful technique for our discrete-time models. 
Recurrence relations %have not previously been used to express coupling dependencies, but they
have been used in related work such as price instability caused by fundamentalist/chartist interaction \citep{Chiarella2006}, clustered volatility caused by feedback effects \citep{Farmer2002},
price instability from time-varying demand \citep{Day1990}, and instability 
from leveraging \citep{Thurner2012}.

Game-theoretic models 
focus on the systemic instability effects of interaction between traders \citep{Giardina2003,Brunnermeier2005}, yet they start from the premise that instability arises from the complexity of traders adapting the value
of some internal parameter in order to optimise a utility function, and the model is used to explore market equilibria.  By contrast, we are interested in systemic effects that arise from traders with fixed strategies where no optimisation is involved and we explore market disequilibria and detailed causation at a microstructure level.

Other studies of feedback in financial markets include:
\citet{Gennotte1990}, who model price instability and show how the extent of a price crash may be determined by the feedback effect arising from unobserved portfolio hedging; 
\citet{Bouchaud1998}, who emphasize the role of feedback effects in market instability (in particular through risk aversion) and utilise a Langevin equation to model feedback mediated via prices; 
and \citet{Westerhoff2003}, whose agent-based numerical simulation explores risk-averse market making strategies in foreign exchange markets and shows how feedback interaction between market makers and speculators can increase trading volume and distort exchange rates.

\citet{Menkveld2014} have recently provided a Markovian recursive model of interaction between HFT market makers and predators; this is in the same spirit as our model, in that it models direct interaction in discrete time and aims to ``uncover effects that remain hidden in static models'' --- however, we develop a more general framework that for example supports a full order book, dependencies on historical values, independent communication delays, and the tracking of multiple variables at each time step.

Our work extends the current understanding of interaction-based instability by examining the 
detailed mechanisms (including latency and feedback) by which HFTs (and others) can become dynamically coupled, causing 
them to operate unintentionally as a collaborative unit that leads to nonlinear oscillation and unstable markets.
We suspect dynamic coupling may be implicated in a range of  previously observed instabilities.
\section{Background to modelling feedback loops}

For our purpose of modelling feedback loops, we consider a market to be comprised of multiple subsystems, which may overlap.
The smallest subsystem is a single component;  components can be any entity 
--- for example, a human trader or a trading algorithm or a news source, though components may be larger (an exchange) or smaller (a risk management subroutine).  A subsystem may 
contain components or further subsystems.
What is important for the model is the \textit{interaction} between these subsystems.

\subsection{Coupling}

We say that if the behaviour of one subsystem influences the behaviour of another 
the latter is \textit{coupled} to the former and the two 
comprise a larger system 
that exhibits \textit{coupling}. 
For example, two HFTs are coupled if one mimics the operations of the other. 
The HFTs would also be coupled 
if one of them acted according to some pattern triggered by the other's 
activity.  
Any two subsystems 
may be coupled 
and
there may exist {\em chains} of coupled subsystems (e.g. one HFT is coupled to another that in turn is coupled to a news source).
If a subsystem is coupled to itself either directly or indirectly via a coupling chain, we define this to be a {\em feedback loop} and all the subsystems in this cyclic chain are said to be {\em mutually coupled}.

We say a coupling is {\em static} if it always persists, with constant strength: 
a coupling is {\em dynamic} if it is transient or has varying strength.
Dynamic coupling is less predictable, hence more dangerous,  than static coupling.
We define market instability as a large change or volatility in one or more market parameters such as market price, traded volume, or frequency of trading.

\subsection{Oscillation, phases and phase-shifting}

\begin{figure}[ht]
~~~~~~~~~\includegraphics[width=0.33\columnwidth]{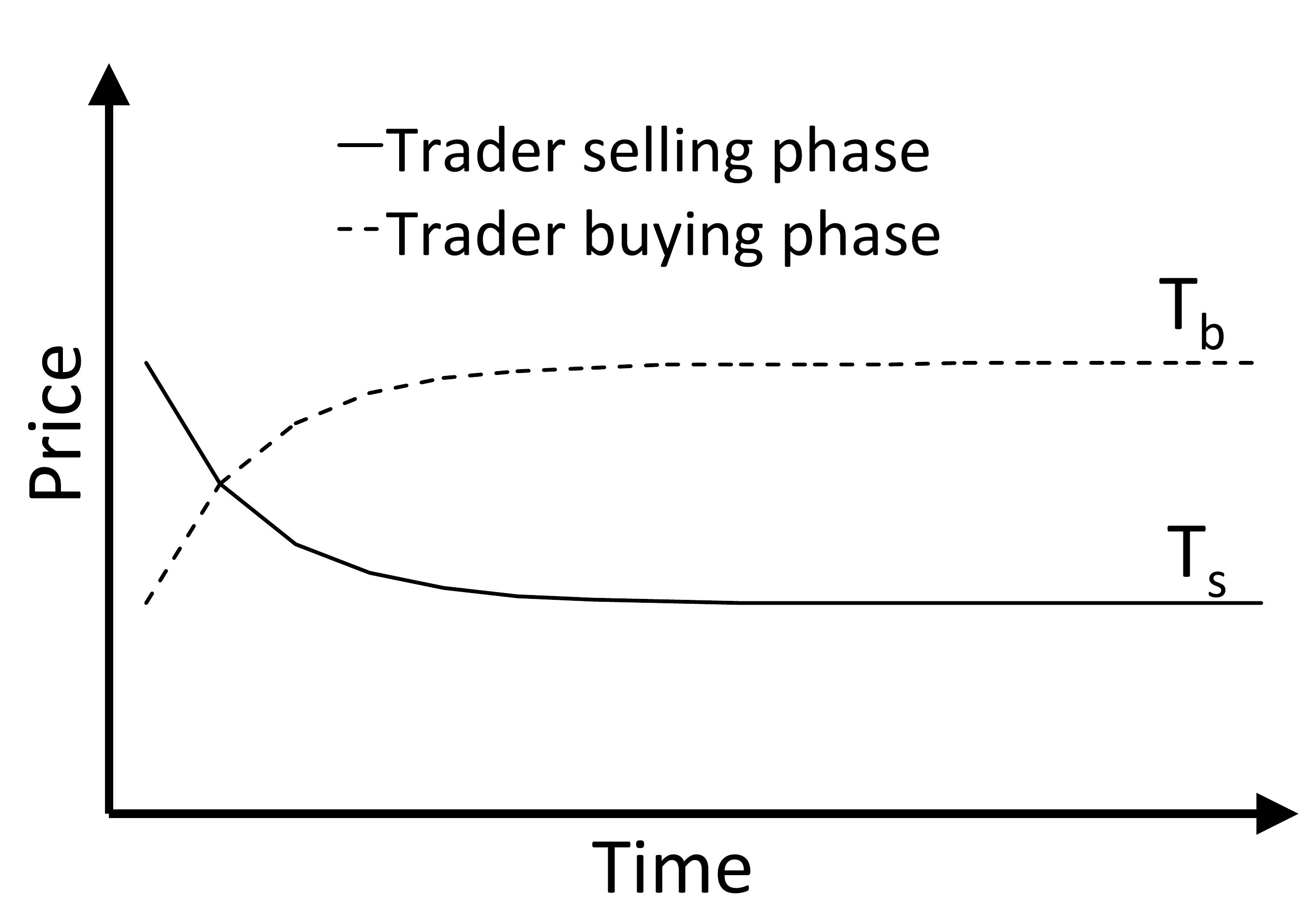}
~~~~~~~~~~~\includegraphics[width=0.33\columnwidth]{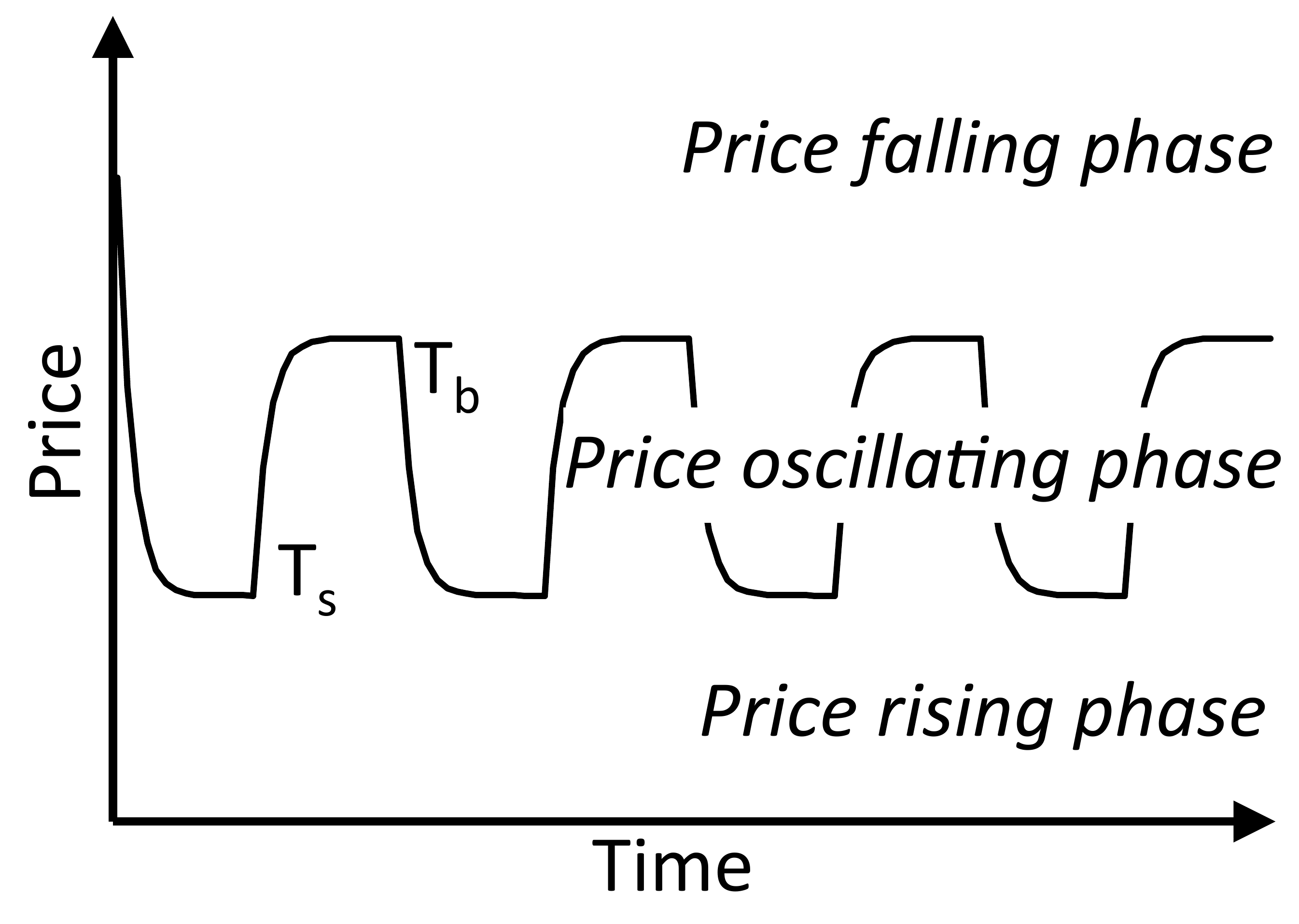}
\caption{Oscillatory instability arising from two stabilising feedback loops: (a) the behaviours of the two trader feedback loops (the lines plot price during trader selling phase and buying phase);  (b) the three phases for the market as a whole --- price bands identified for rising ($price < T_s$), 
oscillating ($T_s \leq price \leq T_b$) 
and 
falling ($price > T_b$) 
prices.}
\label{fig:oscillatingLoop2}
\end{figure}

Oscillation
can arise in many ways: for example, the interaction between momentum  and fundamental traders can lead to oscillatory price behaviour 
\citep{Sethi1996,Chiarella2006}.
Oscillatory instability may also arise from two stabilising feedback loops.  Consider a simple contrarian trader who 
buys when prices are too low, and sells when prices are too high. Assume the prices of the trader's limit orders are set to be halfway between the current market price and a reference price ($T_b$ for buying and $T_s$ for selling); thus the trader's behaviour is \textit{coupled} to the market price.   If these orders affect the market price, causing the price to drop when selling  and to rise when buying, then the market price is also \textit{coupled} to the trader behaviour. 
A two-way coupling is created and each phase (buying/selling) creates a stabilising feedback: when selling, market price asymptotically drops to $T_s$;  when buying, market price asymptotically 
rises to $T_b$. 
Now if
$T_b > T_s$ 
and 
if
the strategy exhibits inertia 
between 
$T_b$ and $T_s$ 
(if it was previously buying(selling), it continues until $T_b$($T_s$), at which point it switches to selling(buying)), the combined effect is an oscillatory instability as illustrated in Figure~\ref{fig:oscillatingLoop2}.

Modelling such a market is complex 
because the behaviour of the market in the region $T_b > P > T_s$ is not entirely determined by the price $P$ but also by the current 
behavioural phase of the trading algorithm.  
The trader in this example displays two behavioural \textit{phases}: buying and selling. 
We also say that the overall market exhibits three phases: a falling-price phase ($P \geq T_b$), a rising-price phase ($P \leq T_s$), and an oscillating-price phase ($T_b > P > T_s$).

More precisely, we say that a \textit{phase} of a subsystem (or of a market) is a set of measurable properties which define a distinctly 
different behaviour of the subsystem,
and the transition from one phase to another we call 
\textit{phase-shifting}.   Phase-shifting by itself is not necessarily a sign of instability, however we shall show later how phase-shifting can lead to instability.

\section{The Model}
\label{sec:model}
In this section we explain
how we model dynamic coupling and feedback in a financial market.  We start with some background comments to explain what we wish to model and why we adopt a particular approach; then we illustrate our method with a case study.  Our initial model is a market with two market makers and an exchange; we then add a fundamental seller, and additional unexpected 
communication latency (delay).

\subsection{Background to the model}
\label{subsec:backModel}

Our aim is to model coupling and feedback at a sufficient level of detail to investigate 
how they arise, how they operate, and how they contribute to market instability.
Our technique is deterministic rather than probabilistic, in order to expose precise 
mechanistic causality. For example, the precise timing and interleaving of order
flow may be critical to the analysis of instability arising from HFT interactions.

In general, when modelling a market with complex feedback loops, the subsystem 
behaviours of interest
may not be expressible in  analytic form, they
may depend on local memory (e.g. whether to buy or sell at a price may depend on whether the price has recently been rising or falling), they
may be rugged 
(non-smooth), and they
may exhibit complex inter-relationships.

Here we present our deterministic discrete-time model using mutually-recursive recurrence relations.
A key characteristic of our model is that it
supports the direct expression of coupling 
in the structure 
of the model. 
The recurrence relations describe how the value of a subsystem parameter changes with time.  
Where one relation references
another,
this indicates a dependency or coupling; where the latter also references the former, this indicates a simple
feedback loop.  For example, consider the following 
relations for parameters $X_t$, $Y_t$ and $Z_t$ (for general functions $f()$, $g()$ and $h()$):\footnote{The functions $f()$, $g()$ and $h()$ might express linear or nonlinear couplings.}

\[
\begin{array}{l  l   l}
X_t= f(Y_{(t-1)})  &~~~~~~
Y_t = g(Z_{(t-1)}) &~~~~~~
Z_t = h(Y_{(t-1)})
\end{array}
\]

In the above equations, $X$ is unidirectionally coupled to $Y$ and there is a bidirectional coupling between $Y$ and $Z$.
In a slightly less abstract example of feedback, consider traders issuing sell orders based on a delta-hedging model: the delta-hedging model depends on the price of a stock index; the index depends on the prices of the component stocks; and the prices of stocks depends on the price impact of sell orders arriving from the traders:
\[
\begin{array}{l l}
sellorders_t&= deltahedge(index_{t-1})  \\
marketprice_t &= priceimpact(marketprice_{t-1},sellorders_{t-1}) \\
index_t &= indeximpact(marketprice_{t-1})
\end{array}
\]

It is sometimes more convenient to use a single function to define the behaviour of two or more parameters simultaneously.
For example, parameters $R_t$ and $S_t$ may be defined as the result of some general function $f()$:
\[
\begin{array}{l l}
(R_t, S_t)&= f(R_{(t-1)}, S_{(t-1)})
\end{array}
\]
Where the above style of description is used, the coupling is made explicit inside the function $f()$ (for example, Equation~(\ref{eqn:matching}) in Section~\ref{sec:coupling} uses a function $match()$ whose definition contains the detailed couplings).

We may also wish to explore the behaviour in time of a market parameter $X$; this is 
achieved by animating the aforementioned equations from some starting values (e.g. $X_0$,  $Y_0$) and plotting the sequence $\{X_0, X_1, \ldots X_{final}\}$.  This provides a time series such as that shown in Figure~\ref{fig:oscillatingLoop2}.
\label{xref:timeseries}

\subsection{Case Study}

To illustrate our technique, we provide as a case study the detailed model of dynamic coupling with oscillatory feedback between risk-averse HFT market makers. 
Although HFT market makers typically straddle multiple exchanges \citep{Menkveld2013}, our case study
comprises a single exchange, with a number $maxi$ of market makers each identified by a number $i$; we then add a fundamental seller.  
In our model the exchange manages two 
limit order books 
 --- 
 $bidbook_t$ for bids, and 
 $askbook_t$ for asks. 

Our method permits
order flow to
be modelled with each trader issuing one order at each time step, or many orders.  In our case study, each trader issues up to four orders per time step (one bid, one ask, one sell and one buy),\footnote{Bids and asks are resting (passive) orders. Buys and sells are aggressive executable orders; e.g. immediately executable limit orders or market orders.} and the exchange processes all orders received at one time step before considering orders received at the next step.

\begin{figure}[htb]
\includegraphics[width=\columnwidth]{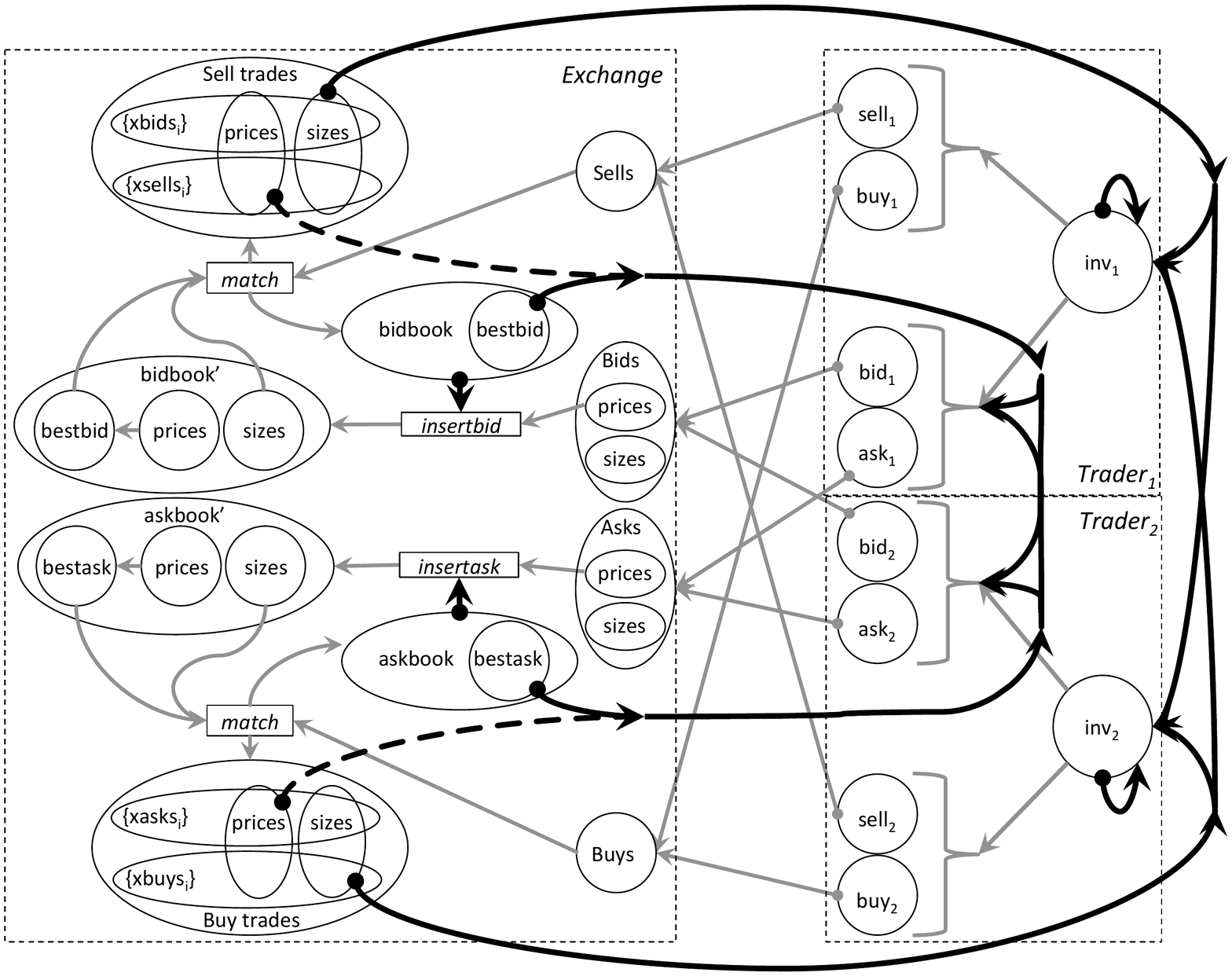}
\caption{Coupling between components for a market with one exchange $Exchange$ and two HFT market makers $Trader_1$ and $Trader_2$.  Arrows are unidirectional dependencies (the head is coupled the tail):  bold black arrows are key feedback dependencies (the dashed arrows are explained in Section~\ref{pricebanding}). If an arrow's tail has an oval it is a time-related dependency --- the head is coupled to the value of the tail at a previous time.
Traders may issue $sell$ or $buy$ orders (determined only by current inventory $inv$) and may also issue $bid$ or $ask$ orders (determined by inventory $inv$ and knowledge of the best bid and best ask at the exchange). Orders are grouped (e.g. $Bids$) before being added to the intermediate $bidbook'$ and $askbook'$ and matched to produce sets of executed orders (e.g. $\{xbids_i\}$, $\{xsells_i\}$) and the resulting $bidbook$ and $askbook$. Each trader only sees his/her own confirmed executions.  Rectangles indicate functions that are described in the text.  In this simple model, crossed bids and asks are not executed against each other; neither are buys against sells.}
\label{fig:coupling}
\end{figure}

\subsubsection{Couplings}
\label{sec:coupling}
Figure~\ref{fig:coupling} illustrates the couplings for a minimal system with an exchange and two 
HFT market makers.
Although this is a simplified version of real market couplings, it is still complex.  
Many 
arrows in Figure~\ref{fig:coupling} represent {\em dynamic} couplings; e.g.
a sell order is determined by the 
difference between a trader's inventory and its inventory threshold, and this varies with time.

Such feedback diagrams 
are not easy to analyse.   We provide a formal, and tractable, description 
by introducing a discrete time dimension.  
Each time ($t$) represents the point when
a message is sent from one entity to another,
and the 
step from $t$ to $t+1$
represents the time taken for an entity
to receive incoming data, 
process it,
and issue a new message. 
Many couplings occur within a time-step but others are {\em time-dependent}  where the
behaviour of one entity is coupled to the behaviour {\em at a previous time} of another entity (in Figure~\ref{fig:coupling} 
these are arrows with ovals at their tails).  Our model requires that every cyclic chain of couplings includes at least one time-dependent coupling.

In the model, all communication occurs synchronously --- either all entities are processing, or they are all sending and receiving messages.
Nevertheless, it is possible to model entities that take differing amounts of time to calculate what to do next (e.g. a slow trader), and it is also
possible to model traders issuing orders at different times.  Both of these effects are achieved by supporting ``empty'' messages; for example, 
a fast trader might send orders  every even timestep and send an ``empty'' message every odd time step, and a slow trader might send orders only on every tenth timestep (otherwise it sends empty messages).  Thus, we can specify the relative speeds of different subsystems.  Furthermore, we shall see in Section~\ref{sec:delays} how
specific delay components can be used to model different latencies in communications links.

For simplicity, 
we assume all traders and the exchange take the same time to process incoming data and issue  messages.  Thus, the total round-trip time 
between a trader and 
the exchange is two time steps, and if a trader waits for a response to one order before issuing the next then that 
trader will only issue orders on alternate time steps. We hereafter assume that traders issue orders on even time steps and the exchange issues confirmations on odd steps.

The value of a parameter at time $t$ may depend on its own previous value at time $t-1$ (or at any previous time, but it may not depend on its value at the same time $t$). This is an example of a time-dependent coupling.
For example, we say that the inventory for trader $i$ at time $t+1$ (denoted by $inv_{i,(t+1)}$) is coupled both to its value at the previous time step $inv_{i,t}$ and to the sizes of the confirmed executions (sent from the exchange at time $t$) of that trader's previously issued orders: $xbids_{i,t}$, $xasks_{i,t}$, $xbuys_{i,t}$ and $xsells_{i,t}$.  

Figure~\ref{fig:coupling} denotes the dependency on the sizes of the executed orders as a time-dependent feedback dependency (coloured black) because the inventory depends on the 
sizes of the executions, which depend on previously issued orders, which in turn depend on previous inventory.\footnote{Although feedback couplings are generally time-dependent, time-dependent couplings need not be feedback couplings.}

We introduce the selection function $\psi(i,x)$ (see \ref{sec:apxA})  to sum the sizes of all (and only) those orders issued by trader $i$ in a set of orders $x$, and our discrete-time recurrence relation for the inventory for market maker $i$ is:
\begin{equation}
\label{eqn:inventory}
inv_{i,(t+1)} = \left \{
\begin{array}{l l l}
inv_{i,t} &+~ \psi(i,xbids_{t}) +  \psi(i,xbuys_{t})& \\
&-~\psi(i,xasks_{t})  - \psi(i,xsells_{t})&
\end{array} \right.
\end{equation}

Equation~(\ref{eqn:inventory}) holds at all time steps, but inventory will only change on even  steps since the confirmations ($xbids$ etc) are only issued on odd steps.

Similarly, we define the couplings (to current inventory $inv_{i,(t+1)}$) that determine the size of each market maker's $sell$ and $buy$ orders.
We introduce the functions $buysize()$ and $sellsize()$ (see Section~\ref{sec:phase-shifting}), which embody the market-maker internal logic for determining buy and sell sizes,  
and the function $order()$, which takes an order type, size, price and identifier, 
and returns an order.
Orders are only issued on even time steps:\footnote{$\nu$ is the price of the executable order --- for the rest of this paper, we assume executable orders are market orders with no price (represented by $\nu$=0).}
\[
\begin{array}{l l}
buy_{i,(t+1)}= order(buy,buysize(inv_{i,(t+1)}),\nu, i)&\\
sell_{i,(t+1)}= order(sell,sellsize(inv_{i,(t+1)}),\nu, i)&
\end{array} 
\]

We also introduce the functions $bidsize()$ and $asksize()$ (see Section~\ref{sec:phase-shifting})
to embody the sizing logic for resting limit order  (coupled to inventory), and the functions $bidprice()$ and $askprice()$ (see below) to embody the pricing logic (coupled to both inventory and order-book information).
This defines how limit orders are coupled to trader inventory and order-book information as illustrated in Figure~\ref{fig:coupling}, where the dependency on order-book information is coloured black to denote a feedback dependency (e.g. because the best bid price depends on the previously issued bids and the previously issued bids depended on the previous best bid price).  Again, for even time steps only:
\[
\begin{array}{l l}
bid_{i,(t+1)}= order(&bid,~bidsize(inv_{i,(t+1)}), \\
&bidprice(bestbid_{t}, bestask_{t}, inv_{i,(t+1)}), i)\\
ask_{i,(t+1)}= order(&ask,~asksize(inv_{i,(t+1)}), \\
&askprice(bestbid_{t}, bestask_{t}, inv_{i,(t+1)}), i)
\end{array} 
\]

\citet{Amihud1980}, \citet{Comerton-Forde2010} and 
\break
\citet{Menkveld2013} show how market makers skew order prices to control their inventories, and
in our case study we use a very simple version of this behaviour: we set limit  order prices such that for high inventory both bid and ask prices are low (encouraging more asks and fewer bids to be executed) and for low inventory both bid and ask prices are high (encouraging more bids and fewer asks to be executed).  
We ensure that bid and ask prices are not negative, new bid prices are not higher than $midprice-1$, and new ask prices are not lower than $midprice+1$ (so resting orders are never crossed). 
\clearpage

Thus in our model the pricing functions have the simple form:\footnote{Detailed expressions for $\alpha$ and $\beta$ are given in \ref{bidpriceaskprice} but are not necessary to understand this presentation.}
\begin{equation}
\label{eqn:bidpriceaskprice}
\begin{array}{l l}
bidprice(bestbid, bestask, inv) &= max (0, (midprice-1) - \alpha\times inv)  \\
askprice(bestbid, bestask, inv) &= max (0, (midprice+1) + \beta\times inv)
\end{array}
\end{equation}

The orders from the two market makers are grouped before being processed by the exchange and we represent these groups as ordered sequences using the notation $\{\ldots\}$ (where $\{\}$ is the empty sequence):
\[
\begin{array}{l}
Bids_{(t+2)} = \{bid_{1,(t+1)} \ldots bid_{maxi,(t+1)}\} \\
Asks_{(t+2)} = \{ask_{1,(t+1)} \ldots ask_{maxi,(t+1)}\} \\
Buys_{(t+2)} = \{buy_{1,(t+1)} \ldots buy_{maxi,(t+1)}\} \\
Sells_{(t+2)} = \{sell_{1,(t+1)} \ldots sell_{maxi,(t+1)}\} \\
\end{array}
\]

The exchange then processes the incoming orders.\footnote{This simple model assumes all orders are guaranteed to be delivered to and accepted by the exchange, though it is also possible to model order confirmations in the case of a system where one or both of these assumptions does not hold.}  
Bids and asks are 
added to the $bidbook$ and the $askbook$ to create intermediate books $bidbook'$ and $askbook'$.
In our model the order books are ordered sequences of limit orders and the positions of the limit orders within an order book are determined by their price and their time of arrival.\footnote{The coupling of $bidbook$ and $askbook$ to order arrival time is not shown in Figure~\ref{fig:coupling}.}
The first bid (ask) in $bidbook'$ ($askbook'$) is the one with the highest (lowest) price (and where there is more than one bid (ask) at that price, they are sorted in order of arrival time so that the earliest arriving bid (ask) is the first in the sequence).  
We introduce the further notational device $x:y$ to represent a sequence of orders where the first order in the sequence is $x$ and $y$ is the remainder of the sequence with $x$ removed.  It follows from the above that if $bidbook = b:bs$ then the best bid is $b$, and if $askbook = a:as$ then the best ask is $a$.
\label{xref:definecolon}

To add new orders to an order book, we introduce the functions 
\break
$insertbid()$ and $insertask()$ (defined in \ref{sec:insa}). 
If we wished bids to rest on the bidbook until cancelled, we would define $bidbook'$ as:
\[
\begin{array}{l}
bidbook'_{(t+2)} = insertbid(bidbook_{(t+1)}, \mathit{Bids}_{(t+2)}) 
\end{array}
\]
However, to simplify the presentation of our case study, we assume that all orders are Fill And Kill (they are fully or partially executed immediately or otherwise cancelled --- also known as Immediate Or Cancel).  We therefore empty the order books before adding newly arrived 
orders:
\begin{equation}
\label{eqn:bidbook}
\begin{array}{l}
bidbook'_{(t+2)} = insertbid(\{\}, \mathit{Bids}_{(t+2)}) \\
askbook'_{(t+2)} = insertask(\{\}, \mathit{Asks}_{(t+2)})
\end{array}
\end{equation}
The exchange then matches the incoming sell (or buy) orders issued at time $t+1$ with those limit orders resting on the $bidbook'_{(t+2)}$ (or $askbook'_{(t+2)}$) to determine the new trade executions $xbids_{(t+2)}$ and $xsells_{(t+2)}$ (or $xasks_{(t+2)}$ and $xbuys_{(t+2)}$) and the new $bidbook_{(t+2)}$  (or $askbook_{(t+2)}$).  
The dependency of $bidbook_{t+2}$ ($askbook_{t+2}$) on $bidbook_{t+1}$ ($askbook_{t+1}$) is another example of a benign feedback and is coloured black in Figure~\ref{fig:coupling}.

The exchange's matching engine is represented by the function $match()$, which must also remove executed limit orders from the relevant order book (discussed below).  Confirmations are only issued on odd time steps:
\begin{equation}
\label{eqn:matching}
\begin{array}{l l l}
(bidbook_{t+2},\ xbids_{t+2},\ xsells_{t+2}) &= match(bidbook'_{t+2},~Sells_{(t+2)}) \\
(askbook_{t+2},\ xasks_{t+2},\ xbuys_{t+2}) &= match(askbook'_{t+2},~Buys_{(t+2)}) \\
\end{array}
\end{equation}

Each executed sell (buy) will be at the price of the currently best bid (ask); and if the size of the sell (buy) is greater than that of the 
best bid (ask), this may change the subsequent best bid (ask) price used for the next execution.  Thus, the executed sell (buy) prices 
are coupled to (i) the current best bid (ask) prices, (ii) the sizes of the executed sell (buy) orders, and (iii) the sizes of the bids (asks) in $bidbook'$ ($askbook'$).  

The operation of the matching engine is complex (see \ref{sec:match}), but we can express the executed sell prices and buy prices inductively using the function $P(r:rr,e:ee)$ where $e:ee$ is a sequence of executable orders to be matched against an ordered sequence $r:rr$ of resting limit orders (using the notation introduced above).  We consider one executable order at a time; we match each executable order against the resting limit orders on the relevant order book --- if there is no liquidity, there are no more trades, but otherwise we have (where $\pi()$ gives the price of an order, $\sigma()$ gives the size of an order, and $\rho(x,y)$ reduces the size of $x$ by the size of $y$):
\begin{equation}
\label{eqn:marketprices}
\begin{array}{l l}
P(r:rr,e:ee)&= \left \{
\begin{array}{l l}
\pi(r):P(\rho(r, e):rr,ee)&if~(\sigma(e) < \sigma(r))\\
\pi(r):P(rr,\rho(e,r):ee)&if~(\sigma(e) > \sigma(r))\\
\pi(r):P(rr,ee)&if~(\sigma(e) = \sigma(r))\\
\end{array} \right.\\
\end{array}
\end{equation}

\begin{figure}[ht]
\begin{center}
\includegraphics[trim = 20mm 65mm 15mm 45mm, clip, width=\columnwidth*5/8]{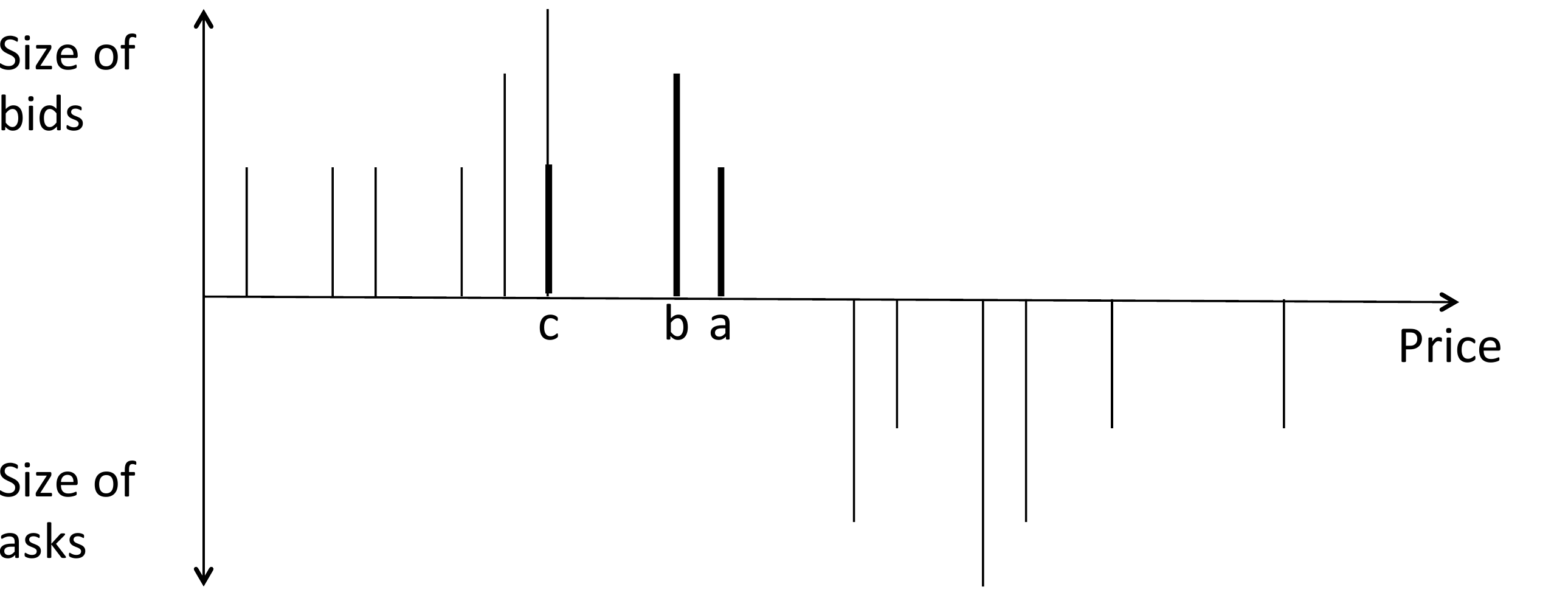}
\caption{Matching sells against bids: the executed prices occur first at the best bid (a); then at (b) since this will have become the new best bid; and finally a larger price jump to (c). The bolded portions of the bid lines are the executions.}
\label{fig:walkingthebook}
\end{center}
\end{figure}

The above equation for market prices specifies exactly how the matching engine ``walks the book'' in order to fill an executable order --- first executing against the best-priced resting limit order and, if that was insufficient to fill the executable order, progressing to the next-best resting order.  
Given a particular distribution of limit orders on the book, a large total size of executable orders is more likely (than a small total size) to deplete the top price level on the relevant order book and cause a jump in the execution price (see Figure~\ref{fig:walkingthebook}).  Whether a price change will occur at all is simply given by comparing the total sizes of all executable orders (in either $Sell$ or $Buy$) with the total sizes of all the resting limit orders resting on the relevant book at the best price (if the former is greater, then at least one execution will be at a different price).  How much the price moves will depend on the distribution of limit orders on the book --- especially the distribution near the top of the book, and the degree of ``gapping'' in that distribution.

The information from $bidbook_t$ and $askbook_t$ is published by the exchange, including the best bid and ask prices which are used by the traders in calculating the next set of limit orders.
The values $xbids_t$, $xasks_t$, $xbuys_t$ and $xsells_t$ are the local trade confirmations sent by the exchange to the relevant traders (the two counterparties to the trade). 
\label{pricebanding}
Finally, Figure~\ref{fig:coupling} contains two bold dashed black arrows which merge with two other feedback arrows --- this represents an optional price banding constraint
that might be applied as a market protection mechanism.\footnote{For a practical example of 
price banding, see \hfill\break http://www.cmegroup.com/confluence/display/EPICSANDBOX/GCC+Price+Banding}

This completes our set of recurrence relations to model the main dependencies illustrated in Figure~\ref{fig:coupling}.
We note in particular that the existence of identified feedback dependencies makes it possible to trace many feedback loops --- including feedback loops that involve both market makers, as 
can be seen from the feedback loops crossing the horizontal midline in Figure~\ref{fig:coupling}.

\subsubsection{Phase shifting}
\label{sec:phase-shifting}
The previous section described how we model dependencies between system components, but we have not yet described the detailed behaviour of the market-making algorithm.  In particular, we wish to model the phase-shifting of an algorithm between two different types of behaviour.  
Computer trading algorithms are frequently subject to phase-shifting, typically implemented as
conditional branches to choose between different behaviours in different market contexts.  Here
we create an algorithm with a somewhat simplistic shift between two dramatically different behaviours --- in practice, an algorithm might exhibit many phases and the switching between phases might be more subtle than this example.

We model a simple risk-averse, long-short market maker that actively manages risk based on the size of the current inventory \citep{Manaster1996}.  Although in practice a market maker could make complex risk calculations, it suffices for our model simply to use raw inventory (since we are only interested in the {\em switching} between behaviours and not precise values).  Our market maker uses a threshold policy \citep{Huang2012} with an upper-bound inventory limit $UL$ and a lower-bound inventory limit $LL$ (a negative number).
To simplify the presentation, these limits are assumed to be fixed, though in practice they 
could vary according to market risk factors such as observed volatility.
Based on these inventory limits, our risk-averse market maker phase-shifts between two different behaviours:
\begin{enumerate}
\item
A stable phase whenever $LL < inv_{i,t} < UL$, where only resting limit orders are issued --- at each even time step (to allow for round-trip communication with the exchange) both a bid and an ask are issued.
We assume that all orders are Fill And Kill.\footnote{See \citep{Chakraborty2011} for a similar model, though in our simple case study we only place one bid and one ask at each time step.}
A special situation arises for $inv_{i,t}=LL+1$ where only a bid is issued and $inv_{i,t}=UL-1$ where only an ask is issued.
\item
A panic phase whenever $inv_{i,t} \geq UL$ or $inv_{i,t} \leq LL$, where 
at each even time step either a large sell or a large buy is issued in an attempt to revert
inventory to zero.\footnote{This aligns (somewhat simplistically) with the empirical observation of \citep{Kirilenko2010a} that ``HFTs do not accumulate a significant net position and their position tends to quickly revert to a mean of about zero'', and with the Nanex description of HFT behaviour during the Flash Crash: ``they slammed the market with 2,000 or more contracts as fast as they could'' \citep{Nanex2010d}.}  
In our model,  executable order size is either $UL$ for a positive-inventory panic or $-LL$ for a negative-inventory panic --- in practice, the size might also depend on market conditions and constraints, but we find this simple approximation is sufficient for our initial model.
No resting limit orders are issued in a panic phase. 
\end{enumerate}

\begin{figure}[ht]
\center
\includegraphics[trim = 70mm 50mm 80mm 60mm, clip, width=0.5\columnwidth]{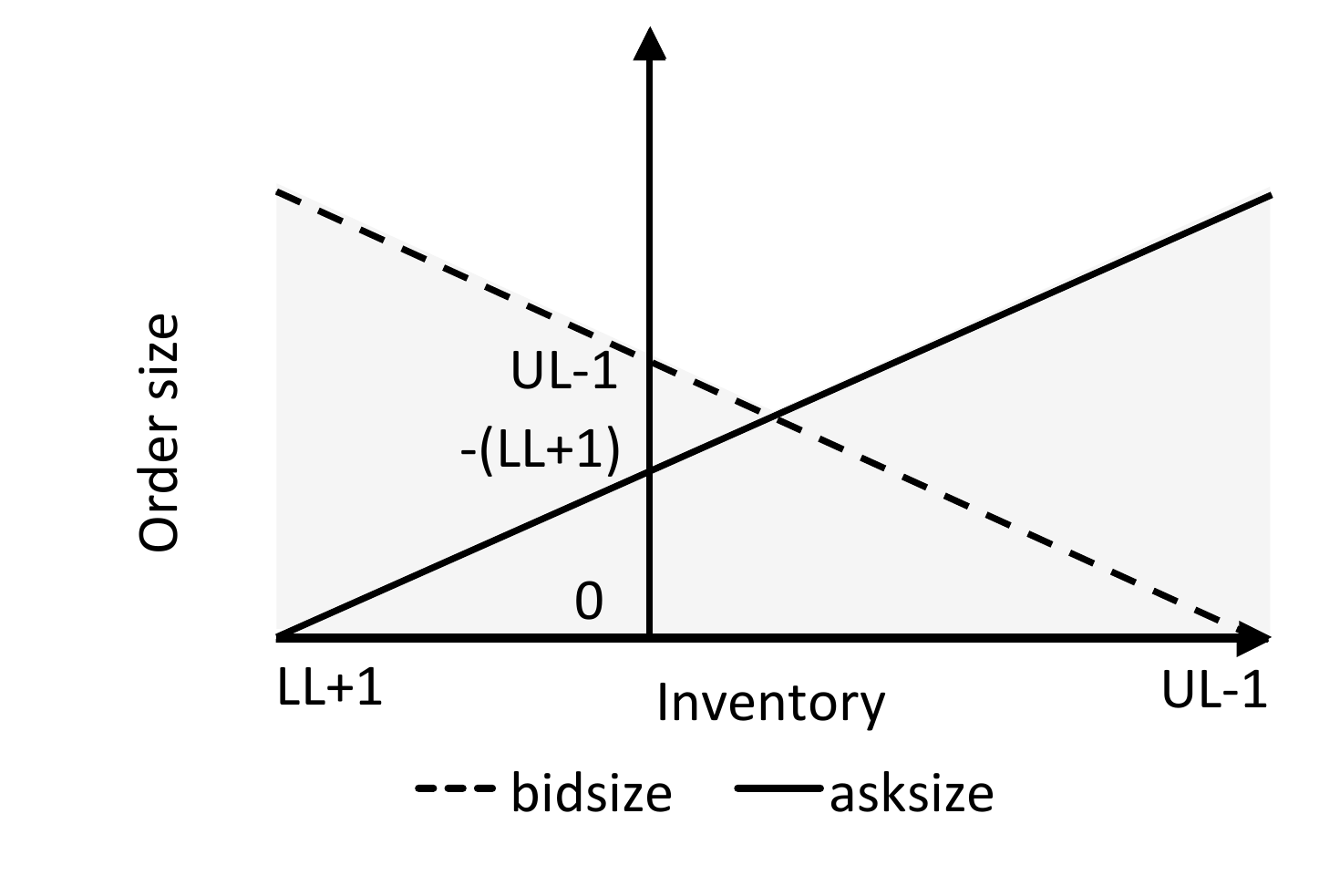}\caption{Risk-averse order sizes as a function of inventory, given inventory limits $LL$ and $UL$. 
Bidsize crosses the y axis at $UL-1$ and asksize crosses the y axis at $-(LL+1)$. 
If bid (ask) size is chosen within the shaded area beneath the ``bidsize'' (``asksize'') line, then it is impossible under normal conditions for inventory to reach the limit 
(see Section~\ref{sec:stability}).}
\label{fig:switching}
\end{figure}

Phase-switching is defined in the functions that determine order size: 
\[
\begin{array}{l}
buysize(inv_{i,t}) = \left \{
\begin{array}{l l}
0&~~~~~~~~~~~~~~if~(inv_{i,t}~>~LL)\\
-LL~~~~~~~~&~~~~~~~~~~~~~~otherwise
\end{array} \right.\\
sellsize(inv_{i,t})= \left \{
\begin{array}{l l}
0&~~~~~~~~~~~~~~if~(inv_{i,t}~<~UL)\\
UL~~~~~~~~~&~~~~~~~~~~~~~~otherwise
\end{array} \right.\\
\end{array}
\]
\[
\begin{array}{l}
bidsize(inv_{i,t})~= \left \{
\begin{array}{l l}
0&if~(inv_{i,t}~\geq~UL)\\
bid size'(inv_{i,t})~~~~~~~~~&otherwise
\end{array} \right.\\
asksize(inv_{i,t})= \left \{
\begin{array}{l l}
0&if~(inv_{i,t}~\leq~LL)\\
asksize'(inv_{i,t})~~~~~~~~~&otherwise
\end{array} \right.\\
\end{array}
\]
Precise limit order size is delegated to the functions $bidsize'()$ and $asksize'()$.  Our model does not require any particular values to be chosen, but we observe that if the limit order sizes are chosen to be within the shaded region of Figure~\ref{fig:switching} then under normal circumstances the market maker will not switch into a panic phase if it starts in the stable phase (see Section~\ref{sec:stability}).  For our case study, we set bid and ask sizes to be exactly the maximum that will never exceed the inventory limits $UL$ and $LL$ (to keep the presentation simple, we ignore details such as minimum size constraints imposed by the exchange, and we use a single large order rather than splitting into several smaller orders).  Thus:
\begin{equation}
\label{eqn:bidsizeasksize}
\begin{array}{l l}
bidsize'(inv) &= max(0,UL-1-inv)\\
asksize'(inv) &= max(0,inv-(LL+1))
\end{array}
\end{equation}

\subsubsection{Adding a fundamental seller}
In the next section we will require a fundamental seller to provide sell orders to trade with the market makers. 
We therefore define the equations for a trader
with index $0$ whose behaviour is to issue a sell order of a fixed size $\omega$ at every even
time step (it does nothing else) up to a predetermined time $timelimit$, and then exits the market.   
Thus (for even time steps only):

\[
\begin{array}{l l l}
buy_{0,t}&= order(buy,0,0, 0) &\\
sell_{0,t}&= \left \{
\begin{array}{l l}
order(sell,\omega,0, 0)&if~(t < time limit) \\
order(sell,0,0,0)&otherwise\\
\end{array} \right.& \\
bid_{0,t}&= order(bid,0,0, 0)&\\
ask_{0,t}&= order(ask,0,0, 0)&
\end{array} 
\]

\subsection{Modelling information delay}
\label{sec:delays}

Information delay is a known and widespread source of instability in the financial markets \citep{Beja1980,Chiarella1992,CFTC-SEC2010a,Tse2012}.
We define information delay to be an unexpected additional latency in transmitting information;
it may manifest in different ways throughout a financial market, 
and may affect all kinds of information.
For example:
(i)
delays in 
financial and economic news, and consequent delays in relevant information being incorporated into traded prices \citep{Beja1980,Chiarella1992};
(ii)
delays due to exchange throttling;
(iii)
delays due to 
technology infrastructure having switched to a business-continuity site;
(iv)
delays in market data: both direct feeds \citep{Nanex2010c,Informa2011,Levin2012,Eholzer2013} and consolidated feeds \citep{CFTC-SEC2010a,Nanex2010c};
(v)
delays and dropouts in the transmission of {\em any} information due to lost or corrupted messages \citep{Corvil2009}; and
(vi)
delays in {\em any} messages to or from an execution venue due to excessive message traffic exceeding the capacities of inbound and/or outbound queues; typically when the market is under stress, but also potentially due to deliberate ``quote-stuffing'' manipulation by traders \citep{Tse2012}.\footnote{Even specialist high-bandwidth interfaces \citep{Eholzer2013} may suffer from delays when traffic is excessive.  Furthermore, if an exchange  provides information about current delays \citep{Eholzer2013} in a normal message, this information will itself be delayed.} 

The extent of delays can be considerable.  For example,  order processing times at Eurex are normally $0.2ms$--$0.35ms$ but can be delayed by a factor of $10$ under normal business conditions and occasionally by a factor of $200$ \citep{Eholzer2013};
and market data reporting from NYSE to the CQS system during the Flash Crash was delayed by  $5,000ms$ -- $24,000ms$ \citep{Nanex2010c}.

Delayed information can substantially affect trading algorithms, since they will make calculations based on incorrect data.  
For example, a risk-averse market maker operating a two-phase strategy as described above may under-estimate inventory risk and this may lead to a phase-shift from ``normal'' to ``panic'' trading.  
Where all traders are affected by delays then systemic effects such as oscillatory instability may ensue.

Consider the introduction of a delay in the trade-confirmation communications link from the exchange to the traders, and let the confirmations of all executed orders be delayed by an additional $\delta$ time steps where $\delta \in \mathbb{N}$.  
We model the delayed information on executed orders as four separate components 
defined as follows:

\[
\begin{array}{l l}
(dxbids_t, dxasks_t) &= (xbids_{(t-\delta)}, xasks_{(t-\delta)})\\
(dxbuys_t, dxsells_t) &= (xbuys_{(t-\delta)}, xsells_{(t-\delta)})
\end{array}
\]

This delay is inserted into our model by specifying that the market maker uses the delayed versions rather than the undelayed versions of the executed orders.
Equation~(\ref{eqn:inventory}) becomes:
\[
\begin{array}{l l l}
inv_{i,(t+1)} &= inv_{i,t} +~ \psi(i,dxbids_{t}) +  \psi(i,dxbuys_{t})& \\
&~~~~~~~~~~-\psi(i,dxasks_{t})  - \psi(i,dxsells_{t})&
\end{array}
\]

The advantages of modelling delays as components are that:
(i)
delays are made both explicit and precise;
(ii)
the extent of individual delays may easily be modified with a simple localised change in the model; and
(iii)
different delays may be inserted at many different points in the market being modelled --- for example, if confirmations of sells were delayed twice as much as other orders, we could write $dxsells_t = xsells_{(t-2\delta)}$.

\subsection{Summary of modelling with recurrence relations}

The foregoing equations 
define 
our model of the main components of our case study: 
the exchange, the market makers, and a fundamental seller.  We claim that 
this style of definition, using mutually-recursive recurrence relations, has the advantage that the 
multiple interactions between the components are made explicit.  For example:
(i) market maker prices $bidprice$ and $askprice$ are coupled to the best bid and best ask prices published by the exchange  (Equation (\ref{eqn:bidpriceaskprice}));
(ii) market maker inventories are coupled to the executed orders $xbids$, $xask$, $xbuys$ and $xsells$ published by the exchange (Equation~(\ref{eqn:inventory})); and
(iii) the executed orders at the exchange are coupled to the orders received from the traders (Equation~(\ref{eqn:matching})).

This equational style provides a highly expressive medium for the description of coupling effects in financial systems with complex dependencies, and we find it to be very useful during the formulation and discussion of hypotheses.  It can be used at varying levels of abstraction (it is not necessary for all components to be modelled at the same level of detail) and it supports a wide variety of real behaviours, including information delay.
\section{From Coupling to Instability}
\label{sec:analysis}
Here we use our case study to illustrate how we reason about coupling-induced instability in a financial market.  
First we review the feedback loops in Figure~\ref{fig:coupling}, and indicate how instability in trader inventories may lead to instability in market prices.  We show that under normal circumstances our market makers are stable, and then we show how instability can be induced by the introduction of an information delay.  The remainder of the section shows how we analyse feedback effects and market instability.

For a different case study, e.g. with different pricing functions
and strategies, the dynamic interaction model would be different but our {\em reasoning process} in relation to coupling, feedback and instability would be the same.

\subsection{Feedback loops}
Figure~\ref{fig:coupling} gives the bilateral couplings for
our case study with one exchange and two market makers, showing how selected components are coupled.  The bilateral couplings
form chains, and the bolded arrows in the figure show key couplings that turn chains into feedback loops.

Limit order prices are coupled to the best bid and ask prices, which are coupled to the prices of previously-issued limit orders.  This forms a feedback loop; either a trader is coupled to him/herself, or a loop covers both traders. 

The two bold dashed arrows show the effect of an optional price-banding constraint where order prices are deliberately coupled to the last traded price, thereby creating a feedback loop.

Inventories are coupled to the sizes of trades, and the sizes of trades are coupled to the order sizes, which themselves are coupled to the previous inventories.  This creates a dynamic feedback loop (Figure~\ref{fig:coupling}) comprising chains of dynamic couplings.  We will later show how this feedback loop can induce inventory oscillation.

Market price is an attribute of the executed trades --- it is not (absent price-banding) a component of a feedback loop,\footnote{In the presence of price-banding, order prices would be coupled to market price, forming another feedback loop.} though it is coupled to the above feedback loop that connects executed trades to inventories.   As the traders' inventories change, so the limit order book is exposed to changing pressure on traded prices.  
The bid book comes under price pressure as the total sizes of all sell orders exceeds the total sizes of all bids at the best price, and the ask book comes under pressure as the total sizes of all buys exceeds the total sizes of all asks at the best price.  
Unbalanced pressure causes 
the market price to move: balanced pressure leads to
liquidity being depleted at the top of both books, 
more volatile traded prices and
increasing spreads.  

\begin{figure}[ht]
\center
\includegraphics[trim = 30mm 40mm 23mm 35mm, clip, width=\columnwidth*5/8]{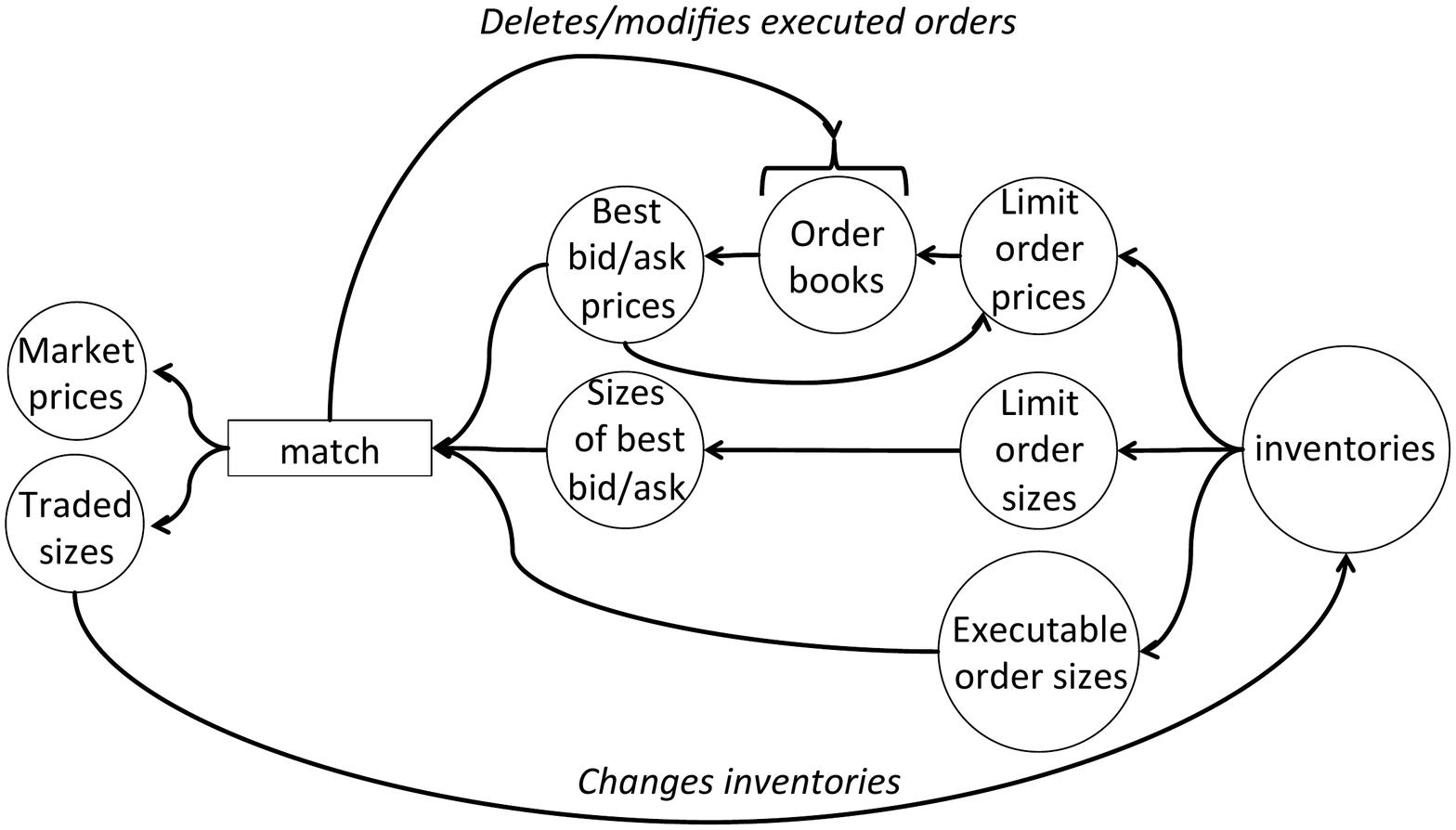}
\caption{How unstable inventories can affect market price.  Order prices and sizes are coupled to inventories, traded prices and sizes are (via the matching function) coupled to order prices and sizes, and inventories are coupled to traded sizes.}
\label{fig:inventorypriceloop}
\end{figure}

Figure~\ref{fig:inventorypriceloop} illustrates the couplings by which unstable inventories might destabilize market price.  The feedback loop between inventories and order sizes can be explored in further detail by expanding Equation (\ref{eqn:inventory}).
The size of each execution is the minimum of the resting order size and the executable order size; at each time step the latter is always either $UL$ for the first sell or $-LL$ for the first buy, and the former is given by Equation (\ref{eqn:bidsizeasksize}):

\[
\begin{array}{l l l}
inv_{i,(t+1)} &= inv_{i,t} &+ \psi(i,xbids_{t}) +  \psi(i,xbuys_{t}) \\
&&- \psi(i,xasks_{t})  - \psi(i,xsells_{t}) \\
&= inv_{i,t} &+ min(UL,bidsize_{i,(t-1)}) +  min(-LL,asksize_{j,(t-1)}) \\
&&- min(-LL,asksize_{i,(t-1)})  - min(UL,bidsize_{j,(t-1)}) \\
&= inv_{i,t} &+ min(UL,max(0,UL-1-inv_{i,(t-1)})) \\
&&+ min(-LL,max(0,inv_{j,(t-1)}-(LL+1))) \\
&&- min(-LL,max(0,inv_{i,(t-1)}-(LL+1))) \\
&&- min(UL,max(0,UL-1-inv_{j,(t-1)}))
\end{array}
\]

The recurrence relation for inventory displays complex feedback:  $inv_{i,(t+1)}$ is not only coupled to its previous values at times {\em t} and {\em t-1} but also to the other trader's inventory at time {\em t-1} ($inv_{j,(t-1)}$).  Furthermore, this recurrence relation only holds if trades occur between the two traders --- yet, if both start in a stable phase there should be no executable orders and no trades.  We shall devote the remainder of this section to the analysis of the dynamic behaviour of our simple market-making strategy: first, to establish its inherent stability, then to demonstrate how it may be destablized, and finally to explore how two or more such market makers may exhibit self-exciting instability.

\subsection{Stability of a single market maker}

Here we analyse the dynamic behaviour of a single market maker's inventory.  
We establish that under normal conditions if our market maker starts in a stable phase it cannot shift 
into a panic phase.  

\label{sec:stability}
A simple algebraic manipulation can be used to establish the stability of the market.  From Section~\ref{sec:phase-shifting}, we know that if $LL+1 \leq inv_{i,0} \leq UL-1$ 
then our market maker issues only resting limit orders.
Consider the extreme case for the maximum achievable inventory --- i.e. when at every time step {\em all} bids and {\em no} asks for the market maker are executed.  Furthermore, recall the prerequisites for our case study ---  that
all limit orders are Fill And Kill (so there are no resting bids from before time {\em t-1}), and that traders issue orders only on even time steps (so $inv_{i,(t-1)}=inv_{i,(t-2)}$ if {\em t} is even).
Now Equation (\ref{eqn:inventory}) may be explored by expanding terms as follows:

\[
\begin{array}{l l l}
inv_{i,t} &= inv_{i,(\textit{t-1})} +~ \psi(i,xbids_{(\textit{t-1})}) +  \psi(i,xbuys_{(\textit{t-1})})& \\
&~~~~~~~~~~~~~-~\psi(i,xasks_{(\textit{t-1})})  - \psi(i,xsells_{(\textit{t-1})})&\\
&= inv_{i,(\textit{t-1})} +~ \psi(i,xbids_{(\textit{t-1})}) - \psi(i,xasks_{(\textit{t-1})}) &\because \text{\footnotesize{only~limit~orders}}\\
&= inv_{i,(\textit{t-1})} +~\psi(i,xbids_{(\textit{t-1})})&\because \text{\footnotesize{only~bids~executed}}\\
&= inv_{i,(\textit{t-1})} +~\psi(i,bidbook'_{(\textit{t-1})})& \because \text{\footnotesize{all~bids~executed}}\\
&= inv_{i,(\textit{t-1})} +~\psi(i,Bids_{(\textit{t-1})})& \because \text{\footnotesize{no~old~bids,~Eq.\ref{eqn:bidbook}}}\\
&= inv_{i,(\textit{t-1})} +~bidsize_{i,(\textit{t-2})}& \\
&= inv_{i,(\textit{t-1})} +~max(0,UL-1-inv_{i,(\textit{t-2})})& \\
&= inv_{i,(\textit{t-1})} +~max(0,UL-1-inv_{i,(\textit{t-1})})&if~t~is~even\\
&= max(inv_{i,(\textit{t-1})}, UL-1)
\end{array}
\]

Thus $(UL-1)$ is a strict, inclusive,
upper-bound for the market maker inventory.  By a similar argument,  $(LL + 1)$ is a strict, 
inclusive, lower-bound. 
We therefore say this market-making algorithm is stable --- it will never reach either of 
its two inventory limits $UL$ or $LL$, and therefore will never panic and will never issue 
aggressive executable orders. 

\subsection{Instability induced by information delay}
\label{sec:infdelay}
If a delay were introduced into the market, unknown to the market maker, such that 
confirmations of all trades were 
delayed by $\delta$ time steps, and if 
there were another trader issuing sells to hit the bids, then we would use the following revised inventory equation, from which (by expanding terms, with the same assumptions as above) we derive a prerequisite for a market maker to shift into panic in such a market.

\[
\begin{array}{l l l}
inv_{i,t} &= inv_{i,(\textit{t-1})} +~ \psi(i,dxbids_{(\textit{t-1})}) +  \psi(i,dxbuys_{(\textit{t-1})})& \\
&~~~~~~~~~~~~~-~\psi(i,dxasks_{(\textit{t-1})})  - \psi(i,dxsells_{(\textit{t-1})})& \\
&= inv_{i,(\text{t-1})} +~\psi(i,dxbids_{(\textit{t-1})})& \\
&= inv_{i,(\textit{t-1})} +~\psi(i,xbids_{(\textit{t-1-}\delta)})& \\
&= inv_{i,(\textit{t-1})} +~\psi(i,bidbook'_{(\textit{t-1-}\delta)})& \because \text{\footnotesize{all~bids~executed}}\\
&= inv_{i,(\textit{t-1})} +~\psi(i,Bids_{(\textit{t-1-}\delta)})& \because \text{\footnotesize{no~old~bids, Eq.\ref{eqn:bidbook}}}\\
&= inv_{i,(\textit{t-1})} +~bidsize_{i,(\textit{t-2-}\delta)}& \\
&= inv_{i,(\textit{t-1})} +~max(0,UL-1-inv_{i,(\textit{t-2-}\delta)})& \\
&= max(inv_{i,(\textit{t-1})}, UL-1+inv_{i,(\textit{t-1})} - inv_{i,(\textit{t-2-}\delta)})
\end{array}
\]

The trader will phase-shift into panic if $inv_t \geq UL$ and from the 
above we therefore have the worst-case pre-condition for shifting to panic that:\footnote{Since traders issue orders only on even time steps, this is equivalent to $inv_{i,(t-2)}>inv_{i,(t-2-\delta)}$ if $t$ is even.}  
\[
inv_{i,(t-1)} > inv_{i,(t-2-\delta)}
\]
 
\subsubsection{Analysing delay (behaviour of shift into panic)}
Consider the case where a market maker's inventory has been stable at value $\nu$ for some time,\footnote{Here, $\nu$ can take any value $\nu<(UL-1)$ --- in Section~\ref{sec:delay2mm} we shall return to this example with $\nu=2-UL$} and then at time $\tau$ a fundamental buyer enters the market and issues very large sell orders at every time step --- sufficient to cause every bid to be executed.  Assume $\delta=2$.  The previously used algebraic manipulation can be applied, and the changes in inventory for the market maker (which only occur on even timesteps) would be:
{\small
\[
\begin{array}{l l|l}
Time&Inventory&Reason\\
\hline
\tau&\nu&xbids_{(\tau-3)}=\{\}\\
\tau+2&\nu&xbids_{(\tau-1)}=\{\}\\
\tau+4&max(\nu,(UL-1)+\nu-\nu)=UL-1&xbids_{(\tau+1)}\neq\{\}\\
\tau+6&max(UL-1,(UL-1)+(UL-1)-\nu)>UL-1&\nu<UL-1\\
\end{array}
\]
}

Thus, the market maker's inventory hits or exceeds the limit $UL$ at time step $\tau+6$, at which point  the market maker shifts into a panic phase and issues executable orders to offload the excess inventory.

\begin{figure}[ht]
\center
\includegraphics[width=0.5\columnwidth]{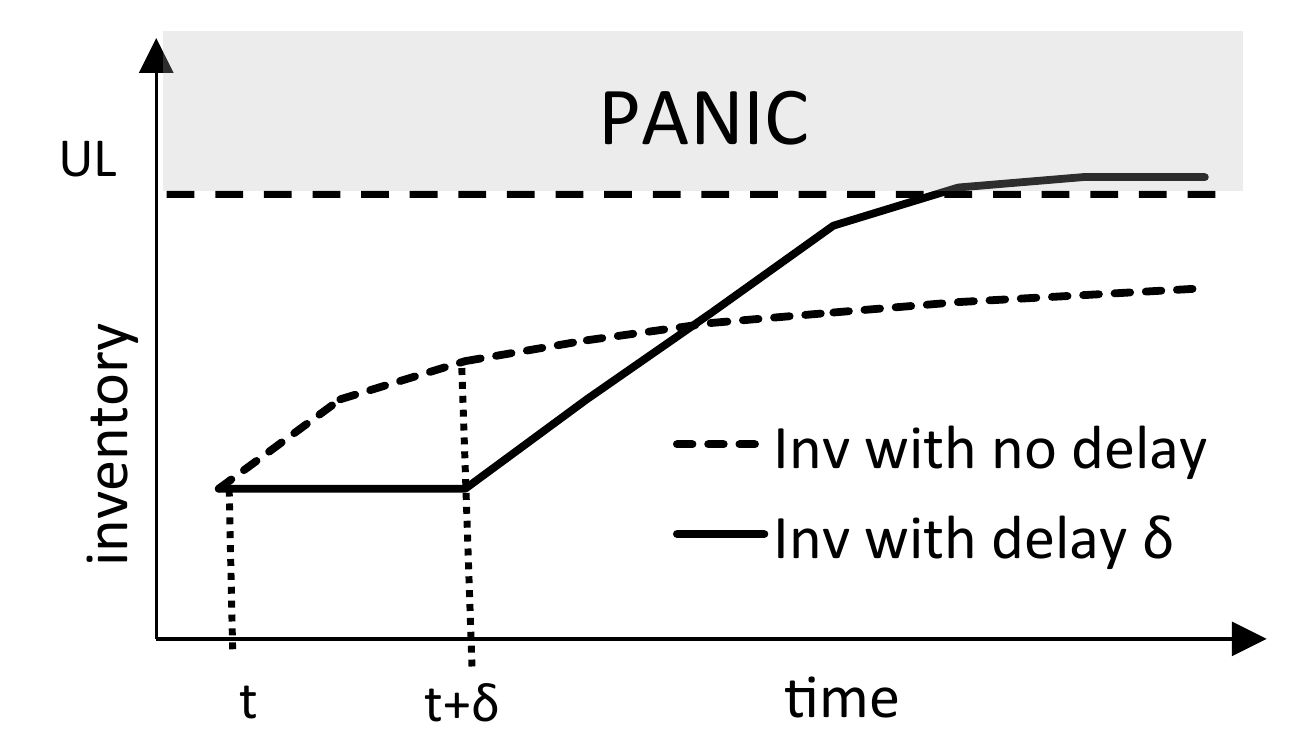}
\caption{Inventory grows with and without delay}
\label{fig:delayWith1mm-Chris}
\end{figure}

More generally, Figure~\ref{fig:delayWith1mm-Chris} illustrates how the introduction of a delay  affects inventory:  without a delay, the inventory asymptotically approaches the limit $UL$ (at a rate that depends on the sizes of its executed bids); whereas if a delay is introduced the inventory initially is unchanged because trade confirmations are buffered in the delay component and the market maker issues more orders based on this unchanged inventory, then the first delayed trade confirmation is received and the inventory increases.  The subsequent increase in inventory is linear for the same length of time as the inventory was previously unchanged (because in our case study the bid sizes and consequently the executions are directly linked to current inventory), and then the inventory increases at a slower rate because the confirmed trades result now from bids issued at higher inventories.   
The inventory then hits or exceeds the limit $UL$ and the market maker shifts into the panic phase.
 
\subsubsection{Analysing delay (shift back into stable phase)}
In the panic phase the market maker will try to return to a stable phase as soon as
possible by issuing a sell order.  Whether this is possible in one transaction
depends on both the extent to which the current inventory exceeds the limit and whether the resting liquidity on the order book is sufficient to fully execute the sell order. To provide such liquidity, thereby permitting the market maker to phase-shift back to a stable phase, we would require the other trader to phase-shift its own behaviour so that it issues bids. In the best case for our case study, $\psi(i,xsells_{(t-1-\delta)}) = UL$ (i.e. the market maker's sell is completely executed) and we have:
\[
\begin{array}{l l l}
inv_{i,t} &= inv_{i,(t-1)} - \psi(i,dxsells_{(t-1)}) & \\
&= inv_{i,(t-1)} - \psi(i,xsells_{(t-1-\delta)})& \\
&=~inv_{i,(t-1)} - UL
\end{array}
\] 

If the inventory is too high ($inv_{i,(t-1)} \geq 2UL$) or the available liquidity is too small ($\psi(i,xsells_{(t-1-\delta)}) \leq inv_{i,(t-1)}-UL$), 
the market maker will stay in panic and  will keep issuing sell orders until (if satisfied) the current inventory falls below the $UL$ limit.  

With the introduction of a very small delay into the market, an oscillatory phase-shifting of another trader can, via unidirectional coupling (with no feedback), induce an oscillatory phase-shifting behaviour in the market maker.   Our equational model illustrates very clearly how this occurs: if the other trader phase-shifts between issuing sells and bids, this leads to a market maker  oscillation between a positive-inventory panic phase and a stable phase as shown above, and by contrast if the other trader phase-shifts between issuing buys and asks, this leads to a market maker oscillation between a negative-inventory panic phase and a stable phase. 

\subsection{Self-exciting feedback with two market makers}

Here we 
analyse a market containing a feedback loop, where
two market makers can be induced into a self-exciting oscillation.
To establish the feedback loop requires a third trader (a fundamental seller to hit the bids), together with a destabilising scenario such as information delay
to send one of the market makers into a panic phase.  
As soon as one of the market makers shifts into panic (it doesn't matter which one, but we assume that they do not both panic at the same time), the third trader is no longer needed and exits the market.  Having achieved a situation where one market maker is in panic and the other is stable,  
they are able to trade with each other; one issues an executable order and the other issues resting limit orders.

The market maker in panic will reduce its inventory by trading with the stable market maker, and this will change the
inventories of both; since the orders subsequently issued by both are dependent on their inventories, there exists a bi-directional coupling between the two market makers.  This creates a feedback loop (involving the two market makers, the exchange and the delay component), and we shall demonstrate how this feedback loop is ``self-exciting'' in that it needs no other component to continue.

This feedback loop can lead to an infinite oscillatory instability between the two market makers, with each
shifting in and out of panic in a synchronised contra-oscillation.  At first such carefully choreographed contra-oscillation
may appear to be unlikely, but our flow analysis will show how the synchronicity arises naturally out of the equations
that describe the market, with the action of one component causing the action of the other component.

\subsubsection{Information delay with two market makers}
\label{sec:delay2mm}
We recreate the delay market described in Section~\ref{sec:infdelay}, but now with two market makers and a fundamental trader, and with a delay $\delta$ in all trade confirmations.
As before, it is a precondition that the fundamental trader leaves the market as soon as one market maker is in panic and the other is stable (if both market makers panic at exactly the same time step, there will be no trades --- the market will remain inactive and therefore stable).

We assume the additional delay $\delta$ from the exchange 
is unknown to the traders and they are unaware that their current inventories may
subsequently be increased or decreased as the result of trade executions that have occurred but
whose confirmations have not yet been received.
Consequently, a stable market maker may issue a limit order that, if executed,
may cause the previously-panicking market maker to become stable and the previously-stable 
market maker to enter a panic phase.

Our model facilitates analysis and understanding of the behaviour of this market,
since it permits the tracking of individual items of the market state (such as orders and confirmations)  
at each time step.  
Table~\ref{tab:DelayMM2} illustrates such detailed flows
--- this flow analysis 
demonstrates how a starting market state at time $\tau$ where market maker $2$ is in positive 
panic (inventory $2UL-2$) and the other is stable (inventory $2-UL$) can \textit{without external impetus} move first to 
a market state where both traders are stable (time $\tau+4$), then to
a state where market maker $2$ is stable and market maker $1$ is in panic (time $\tau+6$), and finally back again to both being stable (time $\tau+10$). 

In this example, the delayed transit of one trade confirmation is highlighted by a succession of three grey cells.\footnote{In general, we need $\delta-1$ {\em pending~xorders} columns for this kind of flow analysis.}
Different patterns of movement in and out of panic are generated with different starting inventories,
except
that there is a precondition that the initial stable inventory is not $UL-1$, since this would result in no bids being 
issued and therefore no trade (the market would be static and stable).

\begin{table}[htb]
\centering
\scriptsize
\begin{tabular}{|l|l|l|l|l|l|l|l|l|}
\hline
time&$inv_1$&$inv_2$&\textit{orders}&\textit{xorders}&\textit{pending}&\textit{dxorders}&\parbox[t]{0.8cm}{$inv_1$}&\parbox[t]{0.9cm}{$inv_2$}\\
&$(t)$&$(t)$&$(size)$&&$xorders$&&$(t+1)$&$(t+1)$\\
\hline
$\tau$&2-UL&\textbf{2UL-2}&\parbox[t]{1.33cm}{$\theta_{b,1,\tau}\\($2UL-3$)$\\$\theta_{S,2,\tau}\\(UL)$}&&&&2-UL&\textbf{2UL-2}\\
\hline
$\tau$+1&2-UL&\textbf{2UL-2}&&\parbox[t]{1.3cm}{($\theta_{b,1,\tau}$,\\$\theta_{S,2,\tau}$)$UL$}&&&2-UL&\textbf{2UL-2}\\
\hline
$\tau$+2&2-UL&\textbf{2UL-2}&\parbox[t]{1.33cm}{$\theta_{b,1,\tau+2}\\($2UL-3$)$\\$\theta_{S,2,\tau+2}\\(UL)$}&&\parbox[t]{1.3cm}{$(\theta_{b,1,\tau},$\\$\theta_{S,2,\tau})$\\$UL$}&&2-UL&\textbf{2UL-2}\\
\hline
$\tau$+3&2-UL&\textbf{2UL-2}&&\cellcolor{LightGray}\parbox[t]{1.3cm}{$(\theta_{b,1,\tau+2},$\\$\theta_{S,2,\tau+2})$\\$UL$}&&\parbox[t]{1.3cm}{$(\theta_{b,1,\tau},$\\$\theta_{S,2,\tau})$\\$UL$}&2&UL-2\\
\hline
$\tau$+4&2&UL-2&\parbox[t]{1.33cm}{$\theta_{b,1,\tau+4}\\($UL-3$)$\\$\theta_{b,2,\tau+4}\\(1)$}&&\cellcolor{LightGray}\parbox[t]{1.3cm}{$(\theta_{b,1,\tau+2},$\\$\theta_{S,2,\tau+2})$\\$UL$}&&2&UL-2\\
\hline
$\tau$+5&2&UL-2&&&&\cellcolor{LightGray}\parbox[t]{1.3cm}{$(\theta_{b,1,\tau+2},$\\$\theta_{S,2,\tau+2})$\\$UL$}&\textbf{UL+2}&-2\\
\hline
$\tau$+6&\textbf{UL+2}&-2&\parbox[t]{1.33cm}{$\theta_{S,1,\tau+6}\\(UL),$\\$\theta_{b,2,\tau+6}\\($UL+1$)$}&&&&\textbf{UL+2}&-2\\
\hline
$\tau$+7&\textbf{UL+2}&-2&&\parbox[t]{1.3cm}{$(\theta_{S,1,\tau+6},$\\$\theta_{b,2,\tau+6})$\\$UL$}&&&\textbf{UL+2}&-2\\
\hline
$\tau$+8&\textbf{UL+2}&-2&\parbox[t]{1.33cm}{$\theta_{S,1,\tau+8}\\(UL),$\\$\theta_{b,2,\tau+8}\\($UL+1$)$}&&\parbox[t]{1.3cm}{$(\theta_{S,1,\tau+6},$\\$\theta_{b,2,\tau+6})$\\$UL$}&&\textbf{UL+2}&-2\\
\hline
$\tau$+9&\textbf{UL+2}&-2&&\parbox[t]{1.3cm}{$(\theta_{S,1,\tau+8},$\\$\theta_{b,2,\tau+8})$\\$UL$}&&\parbox[t]{1.3cm}{$(\theta_{S,1,\tau+6},$\\$\theta_{b,2,\tau+6})$\\$UL$}&2&UL-2\\
\hline
$\tau$+10&2&UL-2&\parbox[t]{1.33cm}{$\theta_{b,1,\tau+10}\\($UL-3$)$\\$\theta_{b,2,\tau+10}\\(1)$}&&\parbox[t]{1.3cm}{$(\theta_{S,1,\tau+8},$\\$\theta_{b,2,\tau+8})$\\$UL$}&&2&UL-2\\
\hline
$\tau$+11&2&UL-2&&&&\parbox[t]{1.3cm}{$(\theta_{S,1,\tau+8},$\\$\theta_{b,2,\tau+8})$\\$UL$}&2-UL&\textbf{2UL-2}\\
\hline
\parbox[t]{0.7cm}{$\tau$+12}&2-UL&\parbox[t]{1.1cm}{\textbf{2UL-2}}&\parbox[t]{1.33cm}{$\theta_{b,1,\tau\text{+12}}\\(\text{2UL-3})$\\$\theta_{S,2,\tau\text{+12}}\\(UL)$}&&&&2-UL&\textbf{2UL-2}\\
\hline
\end{tabular}
\caption{
Inventory and flow analysis for two panicking HFT market makers: 
delay $\delta=2$; time $\tau$ is even; HFTs orders are issued on even timesteps and executed  on odd timesteps. Negative panics (and executable buy orders) never 
occur, so asks are not shown.
Columns 4 to 7 give: orders issued ($orders$);  trades executed ($xorders$);  trade confirmations sent but not yet received ($pending~xorders$); and confirmations received ($dxorders$). Orders are denoted by $\theta_b$ (bid) and 
$\theta_S$ (sell), followed by the size; panic inventories are bolded. One confirmation flow is highlighted in grey.  The rows at times $\tau$ and $\tau$+12 are identical and after time $\tau$+12 the market infinitely repeats the flows and inventories from times $\tau$+1 to $\tau$+12, with both HFTs oscillating in and out of panic.}
\label{tab:DelayMM2}
\end{table}
\clearpage

\subsection{Infinite oscillation}
\label{sec:infosc}
The alternate phase-shifting illustrated in Table~\ref{tab:DelayMM2} is due to the bidirectional
coupling between the two market makers, and this can lead to an infinite oscillation where two market makers
trade with each other indefinitely.  This is highly unusual for market makers, who make a loss on each filled executable order\footnote{In a flat market, they gain the spread on executions of pairs of bid and ask  limit orders, but lose the spread on an executable order.} 
--- this behaviour is not motivated by any economic imperative but is an artefact of the unintentional dynamic coupling between the two automated strategies.
Our analysis demonstrates that an infinite oscillation is theoretically possible by detecting the case where the market state repeats itself  --- in particular, the repetition of a sub-state 
consisting of the two inventories and the outstanding trades whose confirmations have not 
yet been delivered.  This is illustrated in Table~\ref{tab:DelayMM2} at times $\tau$ and $\tau+12$.  
The inventory and flow analysis in Table~\ref{tab:DelayMM2} provides a detailed understanding of how such oscillations may occur, and we have found flow analysis especially helpful in understanding markets with delays. 
The resulting changes in inventory for the two market makers is further illustrated in
Figure~\ref{fig:prob}~(Left).

\label{subsec:multipleTraders}
Manual algebraic manipulation is appropriate for markets involving relatively few instances of coupling, and flow analysis helps to explore market behaviour in great detail over a short timescale.  Although it is possible to automate algebraic manipulation using a symbolic algebra application, we have found numerical simulation to be more helpful for modelling and analysing the behaviour of complex feedback markets over longer timescales; we view numerical simulation as an important component of hypothesis formulation, to assist in clarifying hypotheses and the consequences that ensue from the logic embodied in a given hypothesis given certain initial conditions. 

We have built a numerical simulator (``InterDyne''\footnote{InterDyne was originally coded in the functional language Miranda (\cite{Turner1985},  \cite{Clack1995}).})
that visualizes our model by animating all its 
underlying equations through time (as mentioned in Section~\ref{subsec:backModel}).
This allows us to monitor time-varying interactions between different components. 
Our simulator also allows us to expand our recurrence relations to be substantially more complex and to encompass a much greater range of real behaviour, such as randomised order arrival times at the exchange,
the execution of crossed bids and asks,
and markets with a large number of heterogeneous market makers using different
order pricing
and sizing functions, with different inventory thresholds.  We are therefore able to explore the effect on market instability of different order arrival times, and we are able to 
demonstrate that the emergent market instability is not dependent on a particular choice of market-making strategy (e.g. it is not simply due to resonance between several identical algorithms).

\subsubsection{The phase-shift into panic}
\begin{figure}[ht]
\center
\includegraphics[width=0.43\columnwidth]{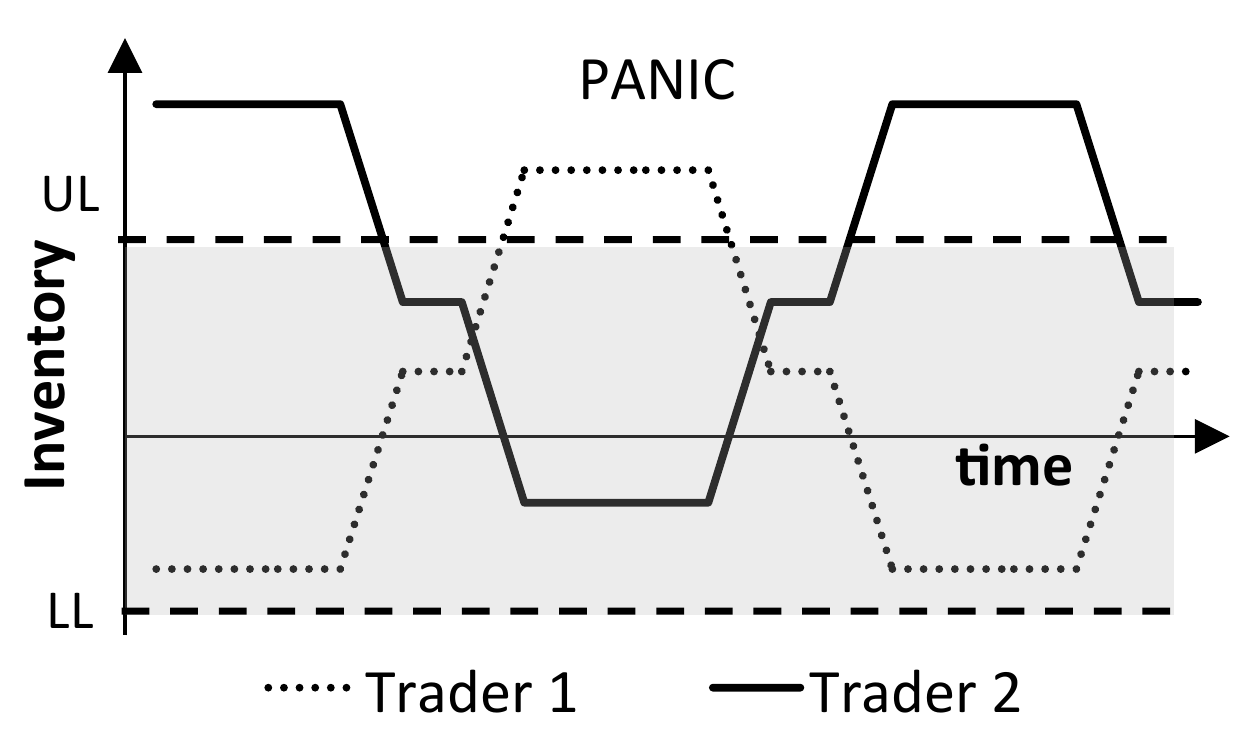}
\includegraphics[trim = 0mm 5mm 5mm 5mm, clip, width=0.45\columnwidth]{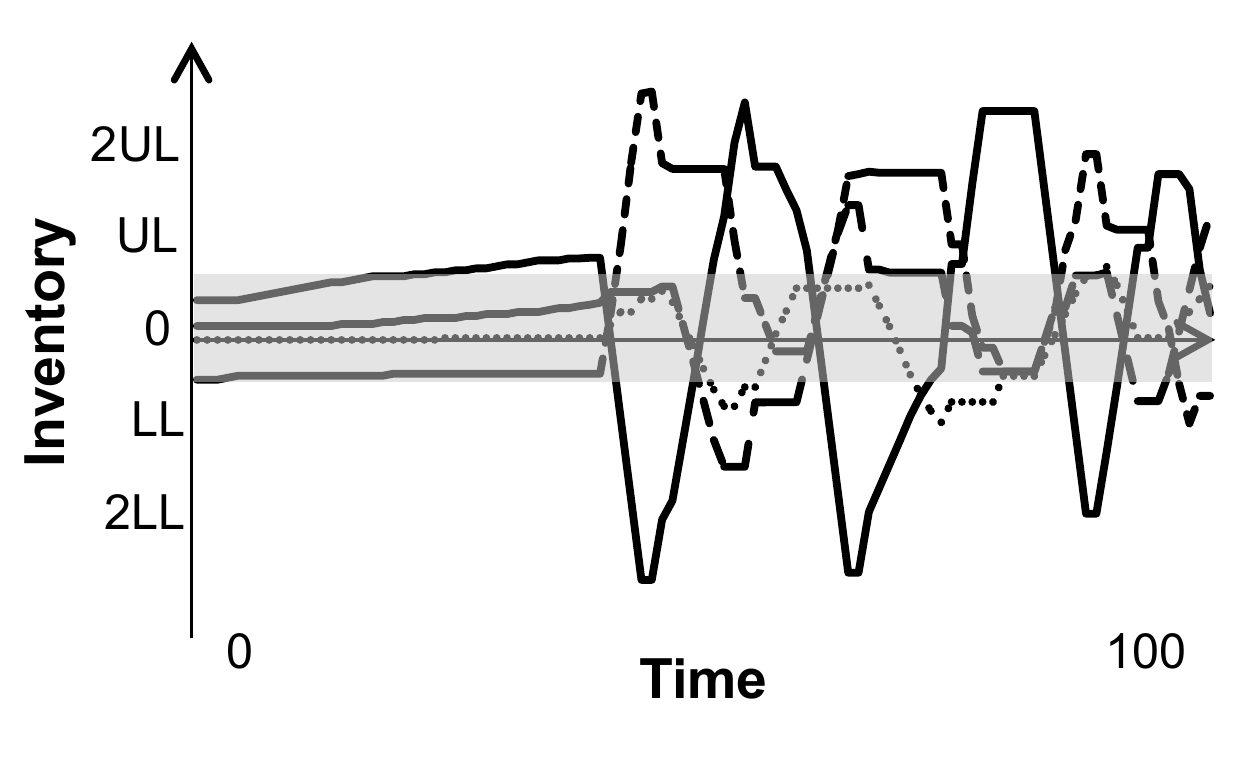}
\caption{Left: oscillatory instability with two homogeneous traders moving in and out of panic in contra-correlation.  Right: oscillating inventories of four coupled heterogeneous traders in a delayed market --- inventories are initially stable, but the market makers are provoked into panic by a single fundamental seller
(not shown) issuing sells at the start of the simulation; this seller  
stops once panic has been provoked.
In both figures the shaded zone indicates the stable region of the trader with the smallest inventory limit.}
\label{fig:prob}
\end{figure}

A necessary precondition for oscillatory instability is that at least one market maker should be in a panic state. We have previously demonstrated how a fundamental trader can provoke a market maker into a panic state if there is an information delay, and Figure~\ref{fig:prob}~(Right) shows this in simulation for four heterogeneous market makers with randomised order arrival times, leading to a phase-shifting oscillation.

In this example, the four market makers have different inventory limits and different order pricing and sizing functions.
For example, resting order sizes are chosen randomly in a range bounded by $0$ and a maximum given by Equation~(\ref{eqn:bidsizeasksize}).  
They are all initially in a stable state within the inventory limits. However, a fundamental seller (not shown in the figure) who issues a fixed number of executable orders in the first few timesteps can lead one market maker to panic, and then the other market makers. 

\subsubsection{Infinite paired coupling}
\begin{figure}[ht]
\center
\includegraphics[trim = 0mm 0mm 0mm 0mm, clip, width=0.8\columnwidth]{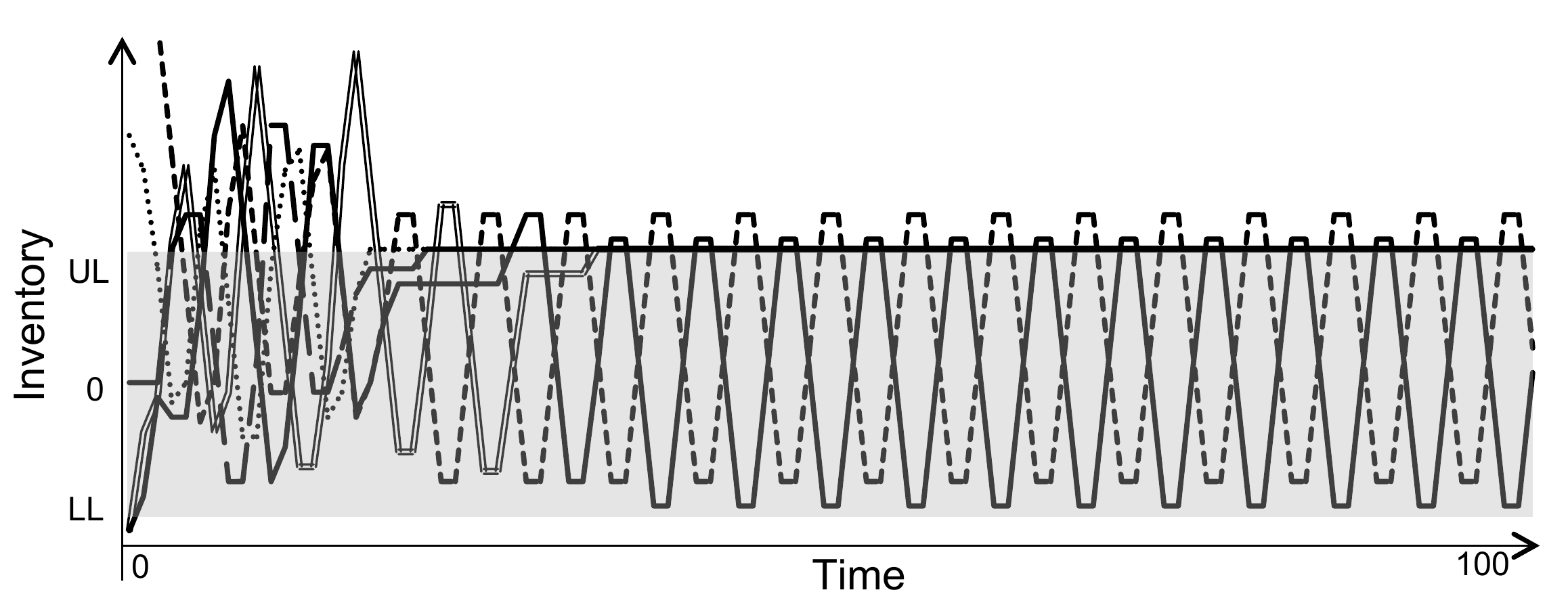}
\caption{Inventory changes for paired coupling in a market with five homogeneous market makers. The shaded zone is a stable state zone within the algorithms' inventory limits.}
\label{fig:paired}
\end{figure}

Figure~\ref{fig:paired} shows the dynamic inventories of five homogeneous market makers  
when a market exhibits a minimal information delay of one time step, and when order arrival
times at the exchange are randomised at each time step.
For this example, two of the traders start trading 
from a positive-inventory panic state, another two traders start from a negative-inventory panic state, and the last trader starts with an inventory of zero.

Figure~\ref{fig:paired} shows the simulation for the first 100 time steps. 
In roughly the first 25 steps all market makers trade among themselves
causing periodic jumps to the panic state and back (due to information delay). 
These jumps are undesired because the executable orders typically incur financial loss; the market makers therefore try to avoid 
those jumps 
by restraining their resting orders when their inventories approach the limits $UL$ and $LL$.    
In the remaining 75 steps three out of five market makers        
manage to stabilize their inventories near the limit  $UL$. 
At an inventory of exactly $UL-1$ they do not issue any bids, and if there are no delayed executions in the pipeline they cannot phase-shift into a positive-inventory panic.
However, the other two market makers remain coupled in a feedback loop and continue to trade with each other (similar to Figure~\ref{fig:prob}~(Left)). This leads to  
an infinite oscillation, where the two market makers repeatedly exchange the same inventory.
This could create a continuously false impression of market
liquidity.

\subsection{How inventory oscillations affect market price}

Figure~\ref{fig:inventorypriceloop} shows how market prices are coupled to the previously described inventory feedback loop, and we have demonstrated how even a very small information delay can trigger that feedback loop to create an oscillating instability in the market maker inventories.  
What we have not yet demonstrated is the extent to which inventory instability can affect market price --- i.e. the strength of the coupling relationship between inventories and prices.

In Section~\ref{sec:coupling} we presented Equation~(\ref{eqn:marketprices}) to specify how the matching engine ``walks the book'' in order to fill an executable order, and 
Figure~\ref{fig:walkingthebook} illustrated how, given a particular distribution of limit orders on the book, a large total size of executable orders is more likely (than a small size) to deplete the top price level on the book and cause a jump in execution price.  

The effects on market price are subtle; different distributions of starting inventories lead to different distributions of orders on $bidbook$ and $askbook$ and therefore different probabilities that a particular executable order will cause a price jump.
However, in our simple case study we found that  a coarse measure of the pressure from large executable orders overwhelming the liquidity on the book can be used as a good ``rough guide'' to changes in price --- it causes prices to change within a single time step, and changes the basic parameters (e.g. best bid and best ask) that drive the pricing functions.

Our coarse measure (for which we make no general claims) subtracts the pressure on resting bids from the pressure on resting asks, and we call this ``Net Liquidity Pressure''; if its value is mostly positive we predict rising prices, and if it is mostly negative we predict falling prices.\footnote{\citet{Menkveld2013} observes
that prices are negatively correlated with HFT inventories.}
In our case study all orders are Fill And Kill, and this  simplifies the definitions enormously:\footnote{The ``$1+$'' in the denominator addresses the case where there is no resting liquidity.}

\[
Net~Liquidity~Pressure_t = \frac{\sum_i\psi(i,Buys_t)}{1+\sum_i\psi(i,Asks_t)} - \frac{\sum_i\psi(i,Sells_t)}{1+\sum_i\psi(i,Bids_t)}
\]

\begin{figure}[ht]
\begin{center}$
\begin{array}{cc}
\includegraphics[width=0.31\columnwidth]{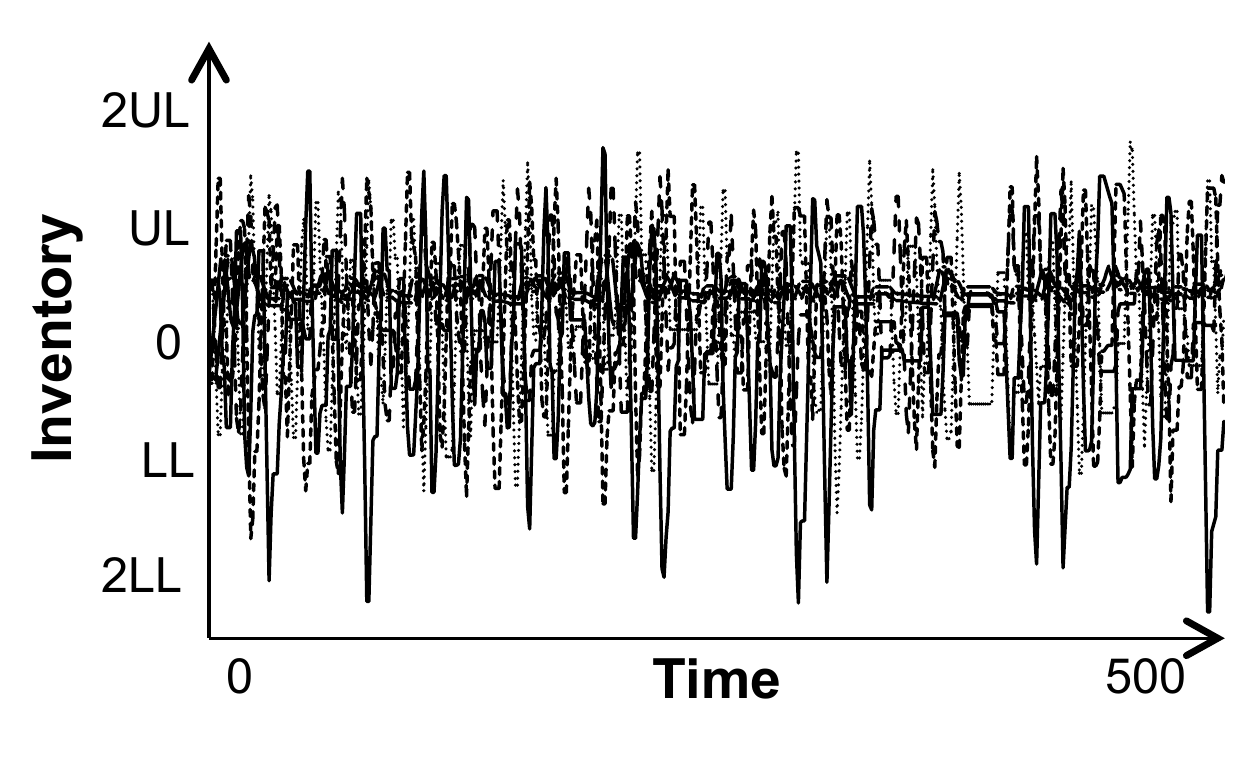}
~\includegraphics[width=0.32\columnwidth]{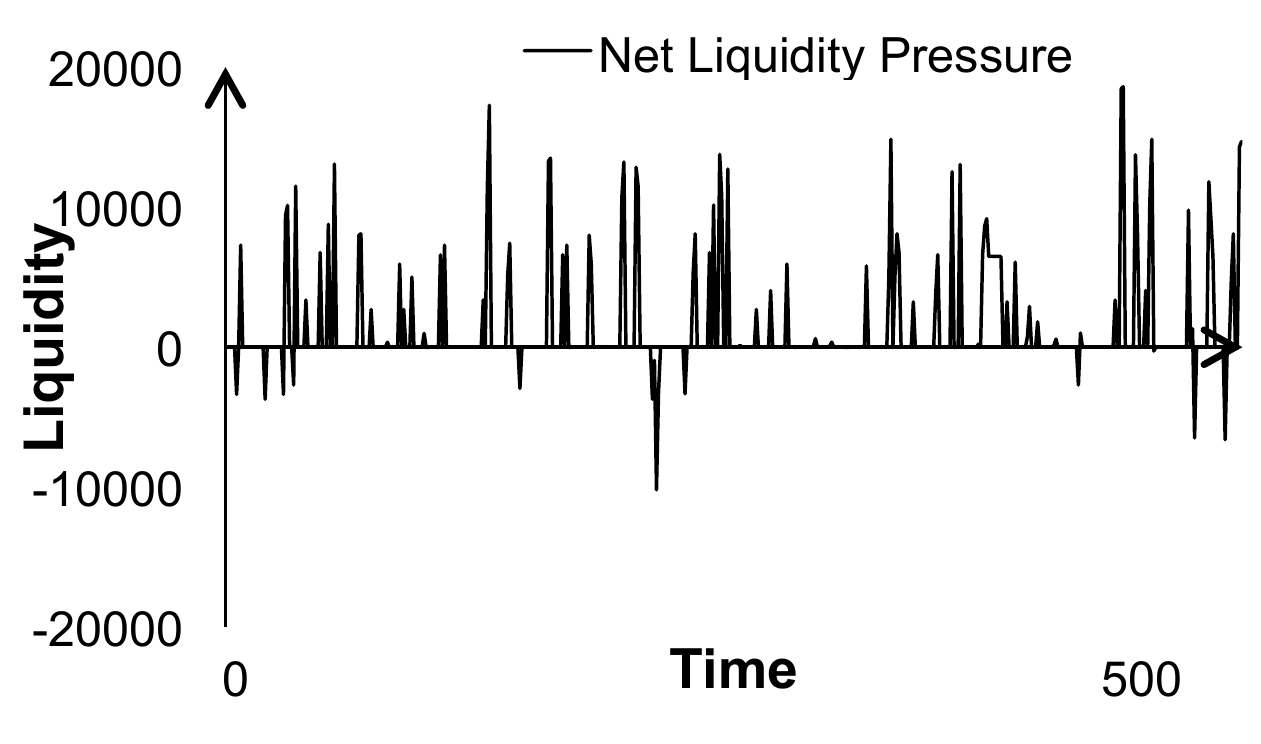}
~\includegraphics[width=0.32\columnwidth]{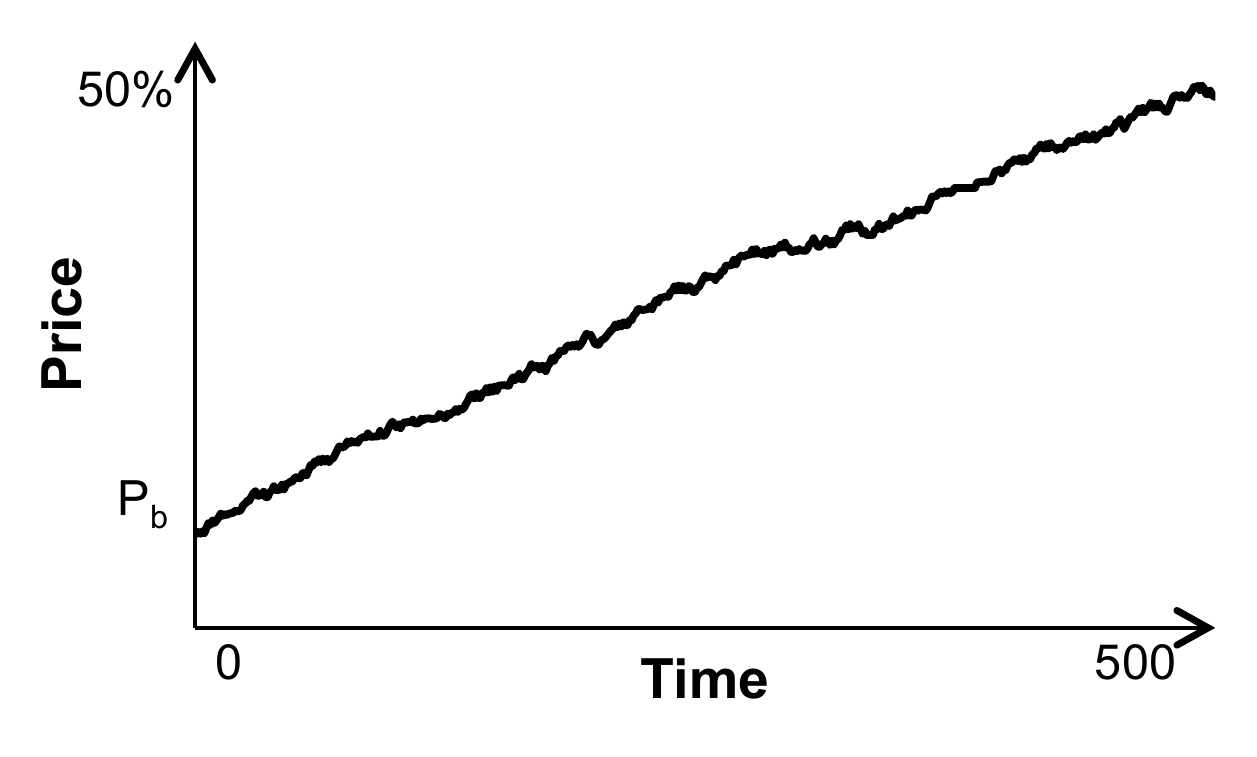}
\end{array}$
\vspace{3mm}

$\begin{array}{cc}
\includegraphics[width=0.31\columnwidth]{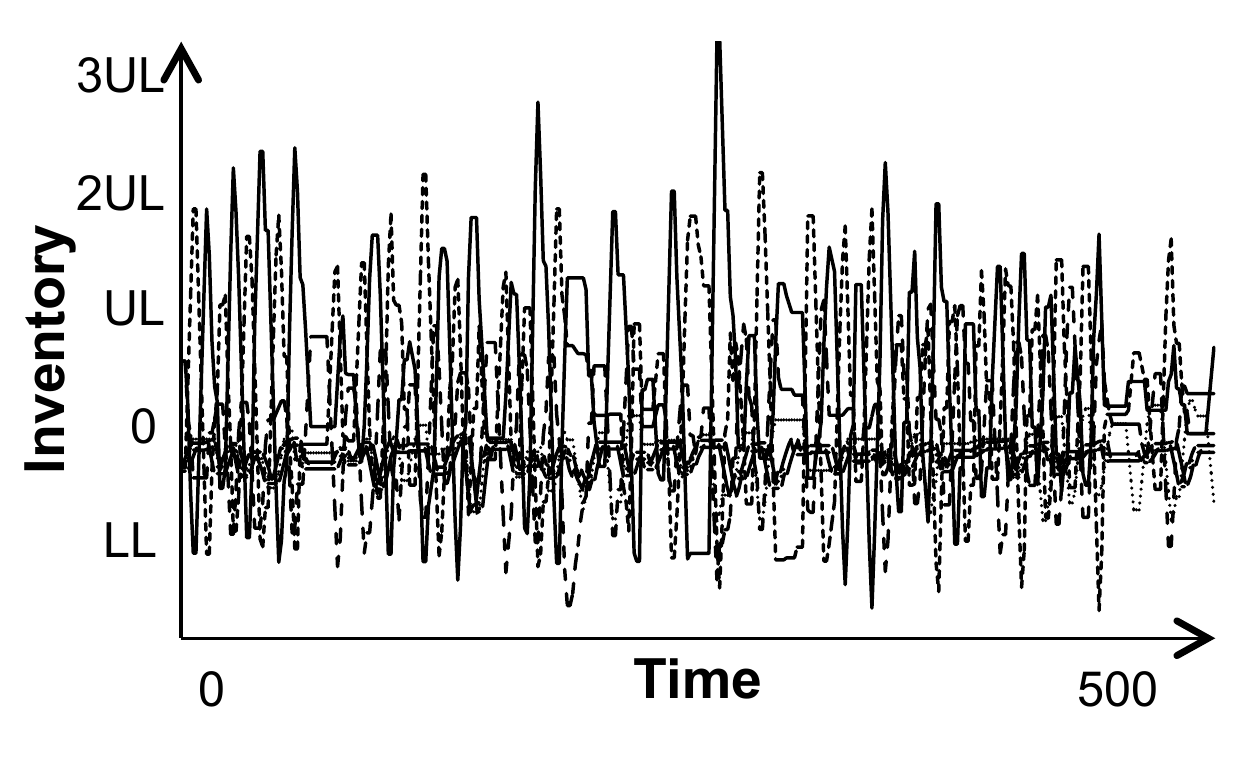}
~\includegraphics[width=0.32\columnwidth]{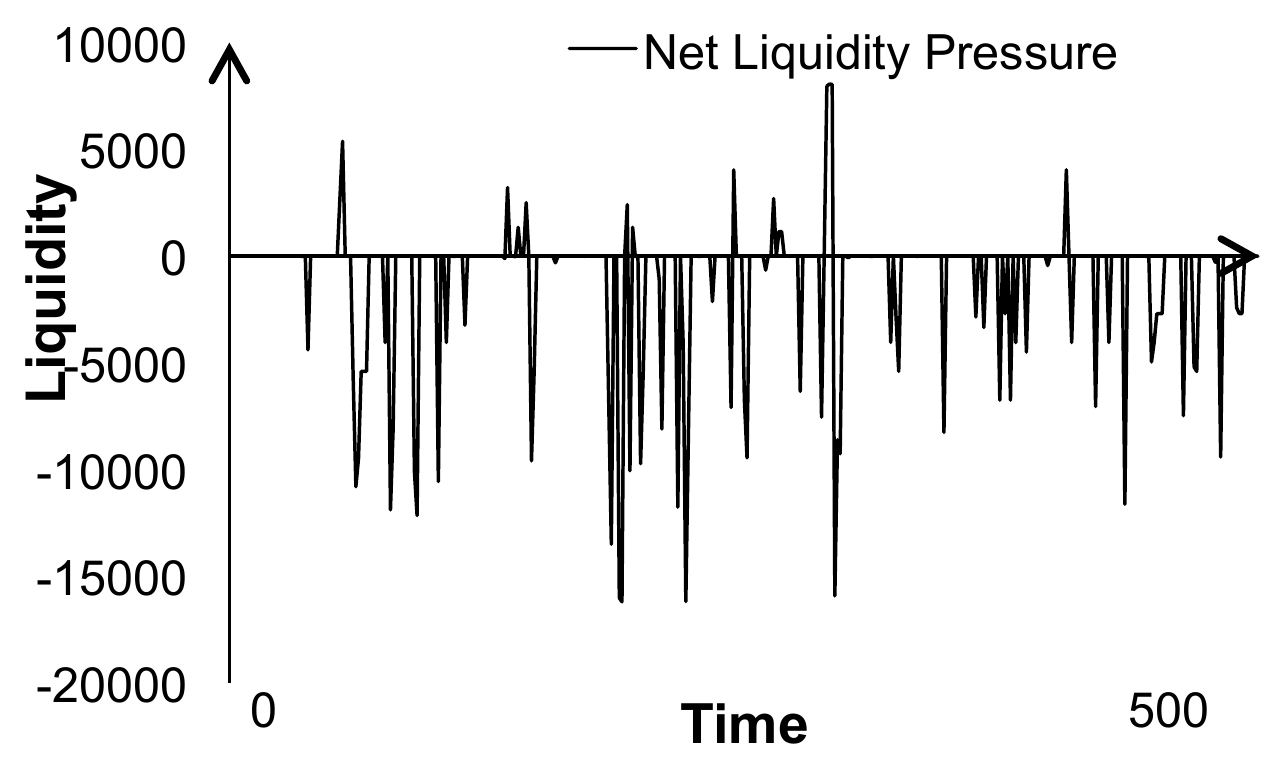}
~\includegraphics[width=0.32\columnwidth]{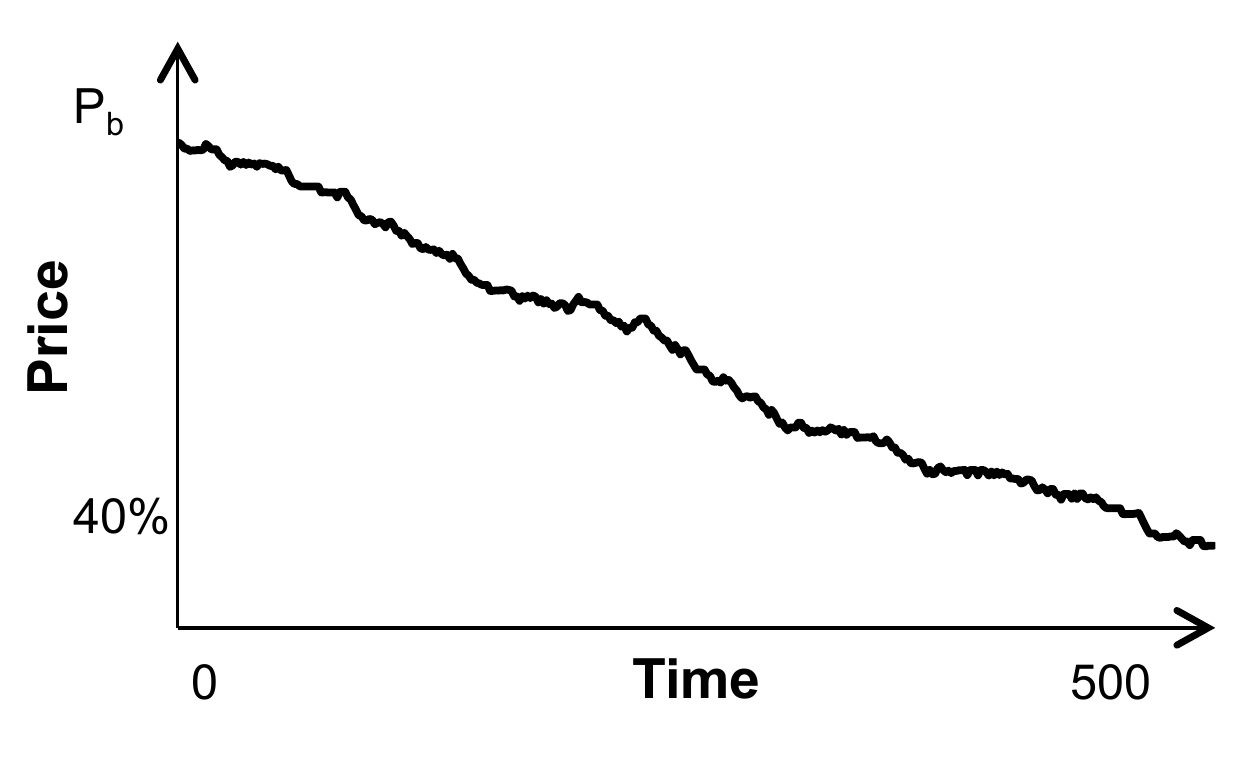}
\end{array}$
\caption{Coupling-induced heterogeneous inventory oscillation, net liquidity pressure, and market prices for two simulations (upper row and lower row).
UL and LL are the limits for the trader with the largest limits and
UL = -LL for all traders. 
}
\label{fig:hetero}
\end{center}
\end{figure}

Figure~\ref{fig:hetero} illustrates the price impact associated with coupling-induced inventory oscillations.  
The results of two numerical simulations are shown, with graphs set out in two rows --- the top row is an example oscillation causing market price to rise, and the bottom row is an example  causing  price to drop.  In each row there are three graphs showing, from left to right, the market maker inventories, the Net Liquidity Pressure, and the market price.

Each simulation comprises a market with an exchange and five heterogeneous market makers (with different inventory limits, different pricing and sizing functions, and where messages to the exchange are randomised at each time step), 
and a delay in trade confirmations of just one time step ($\delta=1$).
In both cases, we assume that prior to the start of the simulation at least one market maker has been induced to panic, and that
the fundamental trader has now withdrawn from the market.  Thus, whatever happens to the price during these
simulations is not due to any fundamental trading --- it can only be due to the trading between the
market makers themselves. 

For the upper simulation, two market makers start with inventories in negative panic and the rest 
have zero inventories: for the lower simulation, two market makers start with inventories in positive panic and the rest have zero inventories.  In both cases, the inventories are 
highly unstable with repeated phase switches into and out of panic (both positive and negative panics).  The net liquidity pressure for the upper simulation is mostly positive, and the rightmost graph shows that market prices rise by about 50\% in 500 time steps (equivalent to about $100ms$); the net liquidity pressure for the lower simulation is mostly negative, and the rightmost graph shows market prices dropping by about 40\% over the same timescale.

\begin{figure}[ht]
\center
\includegraphics[trim = 0mm 0mm 0mm 0mm, clip, width=0.4\columnwidth]{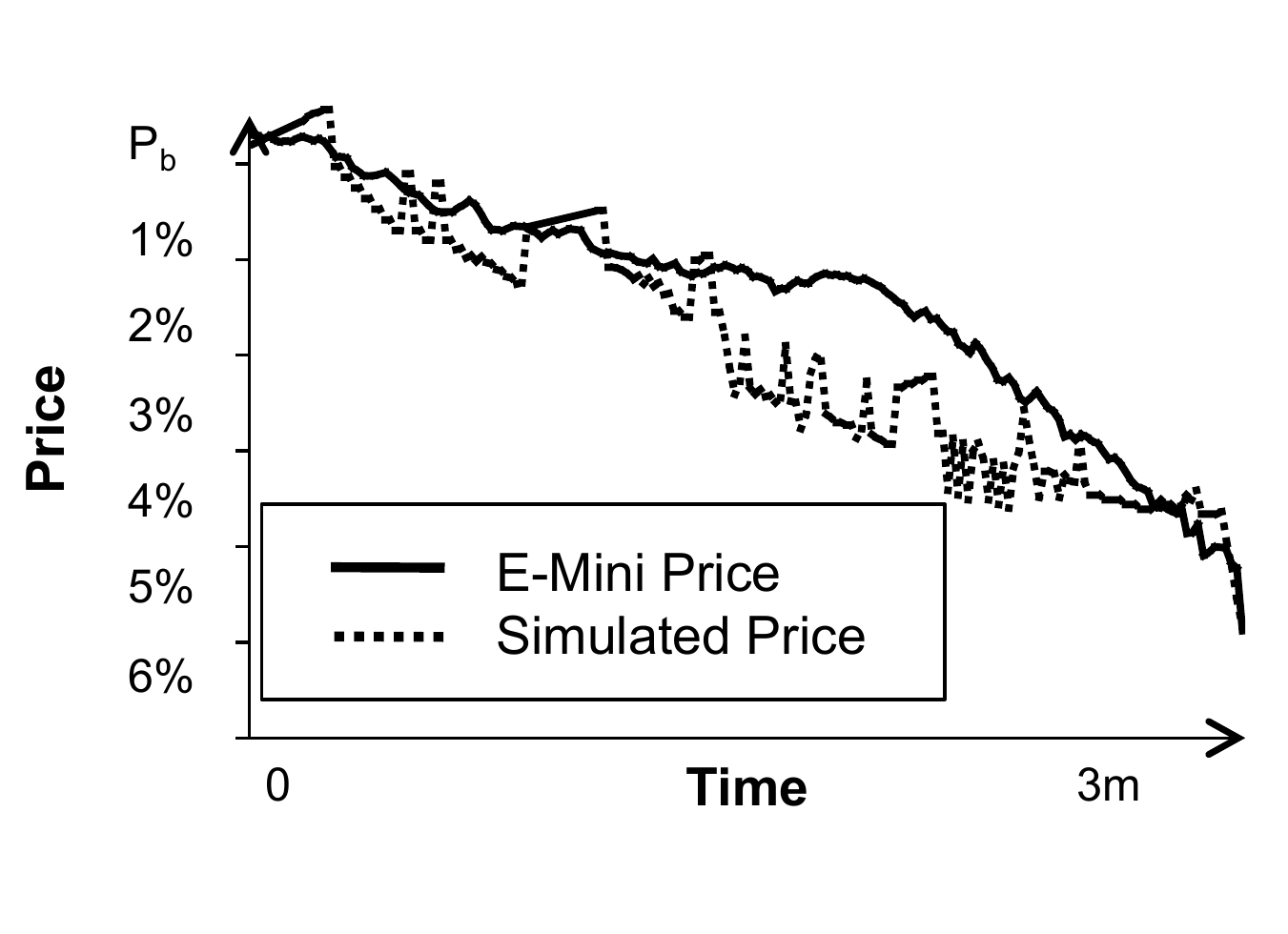}
\caption{Comparison of E-Mini futures price drop during 3 minutes of the Flash Crash (from 14:42.30 to 14:45.27)
with 12 oscillating HFTs, opportunistic and other traders.}
\label{fig:flashCrash}
\end{figure}  

The results of Figure~\ref{fig:hetero} 
indicate 
that if
market makers are induced to trade amongst themselves while other traders
exit from the market, then 
a rapid and appreciable impact on price (up or down) is theoretically possible.

Finally, Figure~\ref{fig:flashCrash} recreates
the price behaviour 
during the Flash Crash of 2010, using public data wherever possible \citep{CFTC-SEC2010a,Kirilenko2010a,Nanex2010e}.
Our simulation uses known factors such as the net HFT inventory, total contracts traded within the selected 3 minutes, and the reported mixture of HFT and Opportunistic traders, but 
there is insufficient public data available for a detailed model and
other factors must be assumed or estimated.  

Even with very limited public data, our model of dynamic coupling and feedback 
provides a reasonable approximation to the key price dynamics.
Figure~\ref{fig:flashCrash} illustrates that it is possible to
use a low-level dynamic interaction model during hypothesis formulation for understanding real events.

Mimicking the price movements of the Flash Crash is not new 
(e.g. \citet{Paddrik2010}), but our approach has the benefit that it is amenable to formal analysis.
We have given some examples of analysis
in this article, and we are  developing more sophisticated techniques.

\section{Conclusion}
\label{sec:summary}

We have demonstrated how coupling between trading algorithms (especially HFTs) can destabilize markets, and have introduced a new technique for modelling dynamic interaction at varying levels of abstraction.  Our case study has shown how unexpected latency and feedback may trigger instability as an unintentional emergent behaviour.

The concept of ``coupling'' (including static, dynamic and time-dependent coupling) has been defined as a bilateral behavioural dependency between subsystems of a market, where a ``subsystem'' has been defined inductively to be a single component or an entity comprising other subsystems.  We have then defined
feedback loops in terms of cyclic chains of couplings, and these definitions underlie our ability to describe a wide variety of market feedback behaviour at multiple levels of detail, and specifically our ability to model dynamic interaction at the level of the market microstructure.

We have introduced a general framework for modelling dynamic interaction and feedback, where recurrence relations in discrete time are used to express the precise nature of bilateral couplings.  Our dynamic interaction models are at a sufficiently low level to express and reason about mechanistic causality, yet are highly flexible in that different parts of the model can be at different levels of detail. The framework also supports the precise expression of communication latencies.  We have demonstrated how such models can be used during hypothesis formulation and can be analysed to provide understanding of the causes and triggers of feedback and prerequisites for instability.  We have also shown how we use numerical simulation to track the time-varying value of a specific variable such as market price, based on a set of starting conditions and a set of recurrence relations to describe a given market; this  provides a further way to analyse the feedback dynamics of a particular model, and we have used this to show how low-level instability in the microstructure of a market can cause high-level instability such as  crashes and spikes in market price.

We have explored unexpected  latency (``delay'') as an example trigger for feedback instability; this has been illustrated with a case study using simple, stable, HFT 
market makers with inventory constraints in an order-book market. We have expained how
dynamic coupling between the HFTs (via the order book) leads to a feedback loop, and 
how delays can then induce
these 
stable 
algorithms 
into an oscillatory instability,
phase-shifting with precisely anti-correlated 
synchrony into and out of inventory panic (``hot potato'' trading).  
We have also shown how coupling-induced feedback between HFTs can be {\em self-exciting} --- in the absence of other effects, it can lead to a theoretically infinite instability.  These effects are induced 
by the size of delay relative to the frequency of trading; thus, 
because short delays occur much more frequently than long delays, HFTs are more likely to suffer from these effects than low-frequency traders.
In broader terms, our analysis suggests that instability can arise as an unintentional emergent behaviour of markets; i.e. it arises not as a consequence of algorithm complexity or predatory behaviour, but instead as a result of transitive interaction effects.  Such emergent instability can arise for a wide range of heretogeneous algorithms with differing order-pricing and order-sizing functions, and is considerably more complex than a 
simple ``resonance'' effect.  
Although we would not expect feedback loops to cause major market instability during equilibrium trading (due to the large mix of strategies \citep{Hasbrouck2013} and because trades within a feedback loop would be outnumbered by other trades), we do expect feedback to become dominant at times of market breakdown when there are fewer traders, and fewer and more correlated trades.

We believe that the concept of feedback as a cyclic chain of bilateral couplings is essential to understanding emergent instability from stable components.
Further, we find that the creation of dynamic interaction models based on recurrence relations is an extremely helpful technique in exploring feedback dynamics, to be used alongside other methods during hypothesis formulation.  
We have not demonstrated how large-scale dynamic interaction models can be constructed and analysed, and clearly there are important issues still to be resolved such as determining 
how to analyse a very large market model to determine whether (and how many) feedback loops exist, to compare the relative importance or strength of different feedback loops, and how likely a given market model is to suffer from feedback-induced instability.

Although our case study focuses on oscillation arising from the interaction of HFT market makers, 
we suspect that many previously observed feedback loops 
\citep{Danielsson2012,Zigrand2012a} may also be modelled and analysed using  our general framework.  Our work may therefore help to 
understand previously unexplained sources of volatility in financial markets; it
may also have implications for models of pricing and market impact, since we demonstrate that traders do not necessarily have independence of action and such models might need to account for unexpected coupling with other traders.

From a practitioner perspective, our dynamic interaction models may help to understand how algorithms and markets could be re-engineered to improve stability.  Since even stable algorithms may be subject to dynamic feedback, traders might now decide to test their algorithms for vulnerability to common modes of feedback instability; execution venues might now decide to offer deterministic latency to improve stability, or to monitor feedback effects and provide enhanced information to subscribers; and regulators might decide to use feedback models to help anticipate the efficacy and consequences of proposed regulation --- especially during periods of disequilibrium, when regulatory control can be particularly important.
%\section*{Acknowledgements}
\appendix
\section{Definitions of functions}
\label{sec:apxA}
\subsection*{\underline{$\psi()$}}

The function $\psi()$  is applied to a sequence of orders $x$ and sums the sizes of all those orders with trader identifier $i$. We use the notation for sequences defined in Section~\ref{xref:definecolon}, and we model each order as a quadruplet $(a,b,c,d)$ containing the type of order ($a$), the size of the order ($b$), the price of a limit order ($c$) and the trader identifier ($d$).
The function $\psi()$  is defined as follows:
{\footnotesize
\[
\psi(i,x) = \left \{
\begin{array} {l l}
\{\}&if~(x=\{\})\\
b + \psi(i,r)&if~(x=(a,b,c,d):r)~and~(d=i) \\
\psi(i,r)&if~(x=(a,b,c,d):r)~and~(d \neq i)
\end{array} \right.
\]
}

\subsection*{\underline{$bidprice()$ and $askprice()$}}
\label{bidpriceaskprice}
$bidprice()$ and $askprice()$ calculate the prices of resting limit orders. The prices are varied linearly according to the current inventory (the aim is that inventory should be zero-reverting).  The functions each take the same three arguments --- the best bid, the best ask, and the inventory.
The bid price is greatest when the inventory is smallest (we set $bidprice = midprice-1$ when inventory is $LL+1$), and the ask price is lowest when inventory is highest (we set $askprice = midprice+1$ when inventory is $UL-1$). We set $bidprice = midprice-1 - \zeta$ when inventory is $UL-1$ and $askprice = midprice+1 + \zeta$ when inventory is $LL+1$, where $\zeta$ is arbitrarily chosen 
(e.g. we use half the CME price band), and we ensure prices do not become 
negative.

{\footnotesize
\[
\begin{array}{l}
bidprice(bb,ba,inv) = max (0, ((ba+bb)/2)-1 - \zeta\times(1-\frac{UL-1-inv}{UL-LL-2}))\\
\\
askprice(bb,ba,inv) = max (0, ((ba+bb)/2)+1 + \zeta\times(\frac{UL-1-inv}{UL-LL-2}))\\
\end{array}
\]
}

\subsection*{\underline{$insertask()$ and insertbid()}}
\label{sec:insa}
\label{sec:insb}
These functions insert a sequence of new orders (argument $z$) into an orderbook sequence (argument $x$).  The orderbook must be sorted to ensure price-time ordering: the first order 
has the lowest price for $askbook$ and the highest price for $bidbook$.  
We use the notation introduced in Section~\ref{xref:definecolon} for sequences.
As explained above, orders are quadruplets --- e.g. $(type,size,price,id)$.  The definition below is for $insertask()$: the definition for $insertbid()$ is identical except that the relational tests are reversed.
{\footnotesize
\[
insertask(x,z)=\left \{
\begin{array}{l l}
x&if~~~(z=\{\})\\
insertask(\{a\},y)&if~~~(x=\{\})\\
&and~(z=a:y)\\
insertask((\tau,\sigma,\pi,i):x,~y)&if~~~(x=(d,e,f,g):q)\\
&and~(z=(\tau,\sigma,\pi,i):y)\\
&and~(\pi < f)\\
(d,e,f,g):(insertask(q,z))&if~~~(x=(d,e,f,g):q)\\
&and~(z=(\tau,\sigma,\pi,i):y)\\
&and~(\pi \geq f)
\end{array} \right. 
\]
}

\subsection*{\underline{match()}}
\label{sec:match}
The function $match()$ 
takes a sequence of limit orders ($l$) and a sequence of market orders ($m$), 
and returns a triple containing 
(i) a revised sequence of limit orders (after executed orders have been deleted, and partial executions amended), 
(ii) a sequence of executed limit orders, and 
(iii) a sequence of executed market orders. 
We either denote an order by a single letter ``$x$'' or by a quadruplet ``$(a,b,c,d)$''
to access its components.
The definition (which discards unmatched market orders) follows the structure of 
Equation \ref{eqn:marketprices}. 

{\footnotesize
\[
match(l,m) = \left \{
\begin{array}{l l}
(\{\},\{\},\{\})&if~~~(l=\{\}) \\
(l,\{\},\{\})&if~~~(m=\{\}) \\
(i,~(a,f,c,d):j,\\
~~~~(e,f,g,h):k)&if~~~(l=(a,b,c,d):q)\\
&and~(m=(e,f,g,h):z)\\
&and~f < b \\
&and~(i,j,k)=\\
&~~~~match((a,b-f,c,d):q,~z) \\
(i,~(a,b,c,d):j,\\
~~~~(e,b,g,h):k)&if~~~(l=(a,b,c,d):q)\\
&and~(m=(e,f,g,h):z)\\
&and~f > b \\
&and~(i,j,k)=\\
&~~~~match(q,(e,f-b,g,h):z) \\
(i,~(a,b,c,d):j,\\
~~~~(e,b,g,h):k)&if~~~(l=(a,b,c,d):q)\\
&and~(m=(e,f,g,h):z)\\
&and~f = b \\
&and~(i,j,k)=match(q,z) \\
\end{array} \right.
\]
}

% References with bibTeX database:

%\pagebreak
\noindent
%{\bf References}
\footnotesize  
\bibliographystyle{elsarticle-harv}
\bibliography{bib}

\begin{thebibliography}{43}
\expandafter\ifx\csname natexlab\endcsname\relax\def\natexlab#1{#1}\fi
\expandafter\ifx\csname url\endcsname\relax
  \def\url#1{\texttt{#1}}\fi
\expandafter\ifx\csname urlprefix\endcsname\relax\def\urlprefix{URL }\fi

\bibitem[{Amihud and Mendelson(1980)}]{Amihud1980}
Amihud, Y., Mendelson, H., 1980. {Dealership Market Market-Making with
  Inventory}. Journal of Financial Economics 8, 31--53.

\bibitem[{Arthur et~al.(1996)Arthur, Holland, Lebaron, Palmer, and
  Tayler}]{Arthur1996}
Arthur, W.~B., Holland, J.~H., Lebaron, B., Palmer, R., Tayler, P., 1996.
  {Asset Pricing Under Endogenous Expectations in an Artificial Stock Market}.
  Tech. Rep. 9625, Social Systems Research Institute, University of
  Wisconsin-Madison.

\bibitem[{Beja and Goldman(1980)}]{Beja1980}
Beja, A., Goldman, M.~B., 1980. {On The Dynamic Behavior of Prices in
  Disequilibrium}. Journal of Finance 35~(2), 235--248.

\bibitem[{Bouchaud and Cont(1998)}]{Bouchaud1998}
Bouchaud, J.-P., Cont, R., Dec. 1998. {A Langevin approach to stock market
  fluctuations and crashes}. The European Physical Journal B 6~(4), 543--550.

\bibitem[{Brock and Hommes(1998)}]{Brock1998}
Brock, W., Hommes, C., Jul. 1998. {Heterogeneous beliefs and routes to chaos in
  a simple asset pricing model}. Journal of Economic Dynamics and Control
  22~(8-9), 1235--1274.

\bibitem[{Brunnermeier and Pedersen(2005)}]{Brunnermeier2005}
Brunnermeier, M.~K., Pedersen, L.~H., 2005. {Predatory Trading}. Journal of
  Finance 60~(4), 1825--1863.

\bibitem[{Caldarelli et~al.(1997)Caldarelli, Arsili, and
  Zhang}]{Caldarelli1997}
Caldarelli, G., Arsili, M., Zhang, Y.-C., 1997. {A prototype model of stock
  exchange}. Europhysics Letters (EPL) 40, 479.

\bibitem[{CFTC-SEC(2010)}]{CFTC-SEC2010a}
CFTC-SEC, 2010. {Findings regarding the market events of May 6, 2010}. Tech.
  Rep. 202, US Commodity Futures Trading Commission \& Securities and Exchange
  Commission.

\bibitem[{Chakraborty and Kearns(2011)}]{Chakraborty2011}
Chakraborty, T., Kearns, M., 2011. {Market making and mean reversion}. In: ACM
  Conference on Electronic Commerce (EC). ACM Press, pp. 307--314.
  
\bibitem[{Chen et~al.(2007)}]{Chen2007}
  Chen, C-C.,  Nagl, S. B.,  Clack, C. D., 2007. {A calculus for multi-level emergent behaviours in component-based systems and simulations}.
  In Proceedings of Emergent Properties in Natural and Artificial Complex Systems (EPNACS 2007), 35--51.

\bibitem[{Chen et~al.(2008)}]{Chen2008} 
Chen, C-C., Nagl, S. B., Clack, C. D., 2008. {A method for validating and discovering associations between multi-level emergent behaviours in agent-based simulations}. In KES International Symposium on Agent and Multi-Agent Systems: Technologies and Applications, Springer. 1--10. 

\bibitem[{Chen et~al.(2009)}]{Chen2009}
  Chen, C-C., Nagl, S. B., Clack, C. D., 2009. {A formalism for multi-level emergent behaviours in designed component-based systems and agent-based simulations}. In From System Complexity to Emergent Properties. Springer. 101--114.

\bibitem[{Chen et~al.(2010)}]{Chen2010}
  Chen, C-C., Clack, C. D., Nagl, S. B., 2010. {Identifying multi-level emergent behaviors in agent-directed simulations using complex event type specifications}. Simulation, 86(1), 41--51.

\bibitem[{Chiarella(1992)}]{Chiarella1992}
Chiarella, C., 1992. {The Dynamics of Speculative Behaviour}. Annals of
  Operations Research 37, 101--123.

\bibitem[{Chiarella et~al.(2006)Chiarella, He, and Hommes}]{Chiarella2006}
Chiarella, C., He, X., Hommes, C., Oct. 2006. {Moving average rules as a source
  of market instability}. Physica A: Statistical Mechanics and its Applications
  370~(1), 12--17.

\bibitem[{Chordia et~al.(2013)Chordia, Goyal, Lehmann, and Saar}]{Chordia2013}
Chordia, T., Goyal, A., Lehmann, B.~N., Saar, G., 2013. Editorial - high
  frequency trading. Journal of Financial Markets 16~(4), 637--645.
  
\bibitem[{Clack(1995)}]{Clack1995}
Clack, C. D., Myers, C., Poon, E., 1995. {Programming with Miranda}.
  Prentice Hall.

\bibitem[{Comerton-Forde et~al.(2010)Comerton-Forde, Hendershott, Jones,
  Moulton, and Seasholes}]{Comerton-Forde2010}
Comerton-Forde, C., Hendershott, T., Jones, C.~M., Moulton, P.~C., Seasholes,
  M.~S., 2010. {Time Variation in Liquidity: The Role of Market-Maker
  Inventories and Revenues}. The Journal of Finance 65~(1), 295--331.

\bibitem[{Cont and Bouchaud(2000)}]{Cont2000}
Cont, R., Bouchaud, J.-P., 2000. {Herd behavior and aggregate fluctuations in
  financial markets}. Macroeconomic Dynamics 4, 170--196.

\bibitem[{Corvil(2009)}]{Corvil2009}
Corvil, 2009. {Low Latency Market Data}. Tech. rep., Corvil Ltd.

\bibitem[{Cvitanic and Kirilenko(2010)}]{Cvitanic2010}
Cvitanic, J., Kirilenko, A., 2010. {High frequency traders and asset prices},
  {W}orking Paper, Cal. Tech.

\bibitem[{Danielsson et~al.(2012)Danielsson, Shin, and
  Zigrand}]{Danielsson2012}
Danielsson, J., Shin, H.~S., Zigrand, J.-P., 2012. Endogenous extreme events
  and the dual role of prices. Annual Review of Economics 4~(1), 111--129.

\bibitem[{Day and Huang(1990)}]{Day1990}
Day, R.~H., Huang, W., 1990. {Bulls, Bears and Market Sheep}. Journal of
  Economic Behavior and Organization 14, 299--329.

\bibitem[{Eholzer(2013)}]{Eholzer2013}
Eholzer, W.~E., 2013. {Some insights into the details that matter for
  high-frequency trading!} Tech. Rep. November, Eurex.

\bibitem[{Farmer and Joshi(2002)}]{Farmer2002}
Farmer, J.~D., Joshi, S., Oct. 2002. {The price dynamics of common trading
  strategies}. Journal of Economic Behavior \& Organization 49~(2), 149--171.

\bibitem[{Gennotte and Leland(1990)}]{Gennotte1990}
Gennotte, G., Leland, H., 1990. {Market liquidity, hedging, and crashes}. The
  American Economic Review 80~(5), 999--1021.

\bibitem[{Giardina and Bouchaud(2003)}]{Giardina2003}
Giardina, I., Bouchaud, J.-P., Jun. 2003. {Volatility clustering in agent based
  market models}. Physica A: Statistical Mechanics and its Applications
  324~(1-2), 6--16.

\bibitem[{Hasbrouck and Saar(2013)}]{Hasbrouck2013}
Hasbrouck, J., Saar, G., Nov. 2013. {Low-latency trading}. Journal of Financial
  Markets 16~(4), 646--679.

\bibitem[{Hommes and Wagener(2009)}]{Hommes2009}
Hommes, C., Wagener, F., 2009. {Complex Evolutionary Systems in Behavioral
  Finance}. In: Handbook of Financial Markets: Dynamics and Evolution.
  Elsevier, pp. 217--276.

\bibitem[{Huang et~al.(2012)Huang, Simchi-Levi, and Song}]{Huang2012}
Huang, K., Simchi-Levi, D., Song, M., Jul. 2012. {Optimal Market-Making with
  Risk Aversion}. Operations Research 60~(3), 541--565.

\bibitem[{Informa(2011)}]{Informa2011}
Informa, 2011. {Sub-zero rate causes Baltic Exchange delay}. [Online; accessed
  27-June-2014] URL
  http://www.lloydslist.com/ll/sector/dry-cargo/article353966.ece.

\bibitem[{Iori(2002)}]{Iori2002b}
Iori, G., Oct. 2002. {A microsimulation of traders activity in the stock
  market: the role of heterogeneity, agents' interactions and trade frictions}.
  Journal of Economic Behavior \& Organization 49~(2), 269--285.

\bibitem[{Kaizoji(2000)}]{Kaizoji2000}
Kaizoji, T., Dec. 2000. {Speculative bubbles and crashes in stock markets: an
  interacting-agent model of speculative activity}. Physica A: Statistical
  Mechanics and its Applications 287~(3-4), 493--506.

\bibitem[{Kirilenko et~al.(2010)Kirilenko, Kyle, Samadi, and
  Tuzun}]{Kirilenko2010a}
Kirilenko, A., Kyle, A.~S., Samadi, M., Tuzun, T., October 2010. {The Flash
  Crash: The Impact of High Frequency Trading on an Electronic Market},
  {R}eport, Univ. Maryland.

\bibitem[{Levin(2012)}]{Levin2012}
Levin, J.~J., 2012. {Mexican Bourse Plans to Resolve Data Delays Before Market
  Opens}. [Online; accessed 27-June-2014] URL
  http://www.bloomberg.com/news/2012-09-14/mexican-bourse-plans-to-resolve-data-delays-before-market-opens.html.

\bibitem[{Lux and Marchesi(1999)}]{Lux1999}
Lux, T., Marchesi, M., 1999. {Scaling and criticality in a stochastic
  multi-agent model of a financial market}. Nature 397~(6719), 498--500.

\bibitem[{Manaster and Mann(1996)}]{Manaster1996}
Manaster, S., Mann, S.~C., 1996. {Life in the Pits: Competitive Market Making
  and Inventory Control}. Review of Financial Studies 9~(3), 953--975.

\bibitem[{Menkveld(2013)}]{Menkveld2013}
Menkveld, A., 2013. {High frequency trading and the new market makers}. Journal
  of Financial Markets 16~(4), 712--740.

\bibitem[{Menkveld and Zoican(2014)}]{Menkveld2014}
Menkveld, A.~J., Zoican, M.~A., 2014. {Need for Speed? Exchange Latency and
  Market Quality}, [Online; accessed 01-July-2014] URL
  http://ssrn.com/abstract=2442690.

\bibitem[{Nanex(2010{\natexlab{a}})}]{Nanex2010e}
Nanex, 2010{\natexlab{a}}. {Flash Crash Summary, Table 2}. Tech. rep., Nanex.

\bibitem[{Nanex(2010{\natexlab{b}})}]{Nanex2010d}
Nanex, 2010{\natexlab{b}}. {May 6th 2010 Flash Crash Analysis Final
  Conclusion}. Tech. rep., Nanex.

\bibitem[{Nanex(2010{\natexlab{c}})}]{Nanex2010c}
Nanex, 2010{\natexlab{c}}. {May 6th Flash Crash Analysis Continuing
  Developments: SEC Report Response}. Tech. rep., Nanex.

\bibitem[{Paddrik et~al.(2012)Paddrik, Hayes, Todd, Yang, Beling, and
  Scherer}]{Paddrik2010}
Paddrik, M., Hayes, R., Todd, A., Yang, S., Beling, P., Scherer, W., 2012. {An
  agent based model of the E-Mini S\&P 500 applied to Flash Crash analysis}.
  In: Computational Intelligence for Financial Engineering \& Economics
  (CIFEr). IEEE, pp. 1--8.

\bibitem[{Sethi(1996)}]{Sethi1996}
Sethi, R., Mar. 1996. {Endogenous regime switching in speculative markets}.
  Structural Change and Economic Dynamics 7~(1), 99--118.

\bibitem[{Thurner et~al.(2012)Thurner, Farmer, and Geanakoplos}]{Thurner2012}
Thurner, S., Farmer, J.~D., Geanakoplos, J., May 2012. {Leverage causes fat
  tails and clustered volatility}. Quantitative Finance 12~(5), 695--707.

\bibitem[{Tse et~al.(2012)Tse, Lin, and Vincent}]{Tse2012}
Tse, J., Lin, X., Vincent, D., 2012. {High Frequency Trading --- Measurement,
  Detection and Response}. Tech. rep., Credit Suisse.

\bibitem[{Turner(1985)}]{Turner1985}
  Turner, D. A., 1985. {Miranda: A non-strict functional language with polymorphic types}. In Proceedings Conference on Functional Programming Languages and Computer Architecture, Springer. 1--16.

\bibitem[{Westerhoff(2003)}]{Westerhoff2003}
Westerhoff, F.~H., 2003. {Market-maker, inventory control and foreign exchange
  dynamics}. Quantitative Finance 3, 363--369.

\bibitem[{Zigrand et~al.(2012)Zigrand, Shin, and Beunza}]{Zigrand2012a}
Zigrand, J.-P.~L., Shin, H.~S., Beunza, D., 2012. {Feedback effects and changes
  in the diversity of trading strategies}. The UK Government Office for
  Science, Foresight Project Driver Review 2.

\end{thebibliography}

\end{document}